\documentclass[12pt]{article}
\usepackage{jheppub}
\usepackage[usenames,dvipsnames]{xcolor}
\usepackage{colortbl}
\usepackage{tikz}
\usetikzlibrary{matrix,arrows,positioning,shapes,snakes,fadings,decorations.pathmorphing,
decorations.pathreplacing,automata,shadows,decorations.markings,patterns,decorations.text}
\usepackage{bm,upgreek}
\usepackage{stmaryrd}

\usepackage{amsmath,amssymb,euscript,array,mathrsfs,appendix,ctable,mathabx,amsthm}
\usepackage{mathtools,float}
\usepackage{arydshln}
\usepackage{todonotes}
\usepackage{graphicx}
\usepackage{wrapfig}
\usepackage{enumitem}
\usepackage{multirow}

\newcommand*\circled[1]{{\footnotesize\tikz[baseline=(char.base)]{%
            \node[shape=circle,fill=black!20,draw,inner sep=1pt] (char) {#1};}}} 
 
\newcommand\SBH{S_\text{BH}}
\newcommand\ZZ{\EuScript S}
 
\usepackage{accents}

\newcommand \arXiv [1]{\href{http://arxiv.org/abs/#1}{\tt arXiv:#1}}

\usepackage{titlesec}
\titleformat{\subsection}[display]{\it}{}{0.1cm}{\vspace{-1.5cm}\begin{center}\thesubsection\hspace{0.2cm}}[\end{center}\vspace{-0.5cm}]

\newcommand{\ket}[1]{|#1\rangle}
\newcommand{\bra}[1]{\langle#1|}
\newcommand{\tr}{\text{tr}}
\newcommand{\ext}{\text{ext}}

\newcommand{\EQ}[1]{\begin{equation}\begin{split} #1
\end{split}\end{equation}}

\title{Grey-body Factors, Irreversibility and Multiple Island Saddles}
\author{Timothy J. Hollowood, S.~Prem Kumar, Andrea Legramandi and Neil Talwar}
\affiliation{Department of Physics, Swansea University, Swansea, SA2 8PP, U.K.}
\emailAdd{t.hollowood@swansea.ac.uk,s.p.kumar@swansea.ac.uk,\\ andrea.legramandi@swansea.ac.uk,n.talwar.2017429@Swansea.ac.uk}
\abstract{
We consider the effect of grey-body factors on the entanglement island prescription for computing the entropy of an arbitrary subset of the Hawking radiation of an evaporating black hole. 
When there is a non-trivial grey-body factor, the modes reflected back into the black hole affect the position of the quantum extremal surfaces at a subleading level with respect to the scrambling time. The grey-body factor allows us to analyse the role of irreversibility in the evaporation. In particular, we show that irreversibility allows multiple saddles to dominate the entropy, rather than just two as expected on the basis of Page's theorem. We show that these multiple saddles can be derived from a  generalization of Page's theorem that involves a nested temporal sequence of unitary averages. We then consider how irreversibility affects the monogamy entanglement problem.
}

\begin{document}

\maketitle

\section{Introduction}

The recognition that certain types of instantons, the `replica wormholes', contribute to entropy of Hawking radiation has finally revealed how semi-classical techniques based on the saddle points of the functional integral can shed light on the information loss paradox without needing a detailed knowledge of the underlying microscopic theory of quantum gravity. The central observable is the von Neumann entropy of a subset $R$ of the Hawking radiation $S(R)$ and the new insight is that there are now two saddles of the gravitational functional integral that can compete to  $S(R)$. One of them leads to the entropy of Hawking's original calculation \cite{Hawking:1974sw,Hawking:1976ra} while the new saddle is given by the contribution of a replica wormhole \cite{Penington:2019kki,Almheiri:2019qdq}.\footnote{See also earlier work 
	\cite{Engelsoy:2016xyb,Almheiri:2019psf,Penington:2019npb,Almheiri:2019yqk} and the review \cite{Almheiri:2020cfm} as well as the related work \cite{Kawabata:2021hac,Bousso:2021sji,Wang:2021woy,Karananas:2020fwx,Hayden:2020vyo,Basak:2020aaa,Colin-Ellerin:2020mva,Goto:2020wnk,Matsuo:2020ypv,Hernandez:2020nem,Ling:2020laa,Chen:2020hmv,Johnson:2020mwi,Chen:2020jvn,Chandrasekaran:2020qtn,Li:2020ceg,Chen:2020uac,Hashimoto:2020cas,Giddings:2020yes,Gautason:2020tmk,Chen:2020wiq,Chen:2019iro,Almheiri:2019hni,Hollowood:2020cou,Hollowood:2020kvk,Hollowood:2021wkw}.} The competition between the two saddles reproduces Page's theorem \cite{Page:1993wv,Page:1993df}, that gives the entropy of one of the factors of a bipartite quantum system ${\cal H}_R\otimes{\cal H}_B$ in some random, or `typical', pure state,
\EQ{
S(R)=\min(\log d_R,\log d_B)\ ,
\label{eq:page_theorem}
}
a result that is valid when $d_R\gg d_B$ or vice-versa. In this case, we identify $\log d_R$ with the entropy of the radiation (suitably regularized) up to a certain time and $\log d_B$ with the Bekenstein-Hawking entropy of the black hole. The former is the Hawking saddle and the latter the replica wormhole one.
At early times the Hawking saddle dominates and the entropy increases as more radiation is produced. Then at late times there is a crossover to the replica wormhole saddle giving an entropy approximately that of the Bekenstein-Hawking entropy of the black hole, which is decreasing. In this way, at least as far as the entropy is concerned, unitary is satisfied.

\begin{figure}[ht]
\begin{center}
\begin{tikzpicture} [scale=0.9]
\draw[fill=Plum!10!white,Plum!10!white] (4.7,7.3) .. controls (4.2,7.8) and (2.5,6.5) .. (1.9,6) to[out=-145,in=40] (1,5.2) to[out=-140,in=0] (0,4.6) -- (0,0) -- (6,6) -- (4.5,7.5);
\draw[fill=yellow!60,yellow!30,opacity=0.6] (0,0) -- (0,7) -- (0.4,7) to[out=-80,in=75] (0,0);
\draw[decorate,very thick,black!60,decoration={zigzag,segment length=1.5mm,amplitude=0.5mm}] (0,7) -- (3,7);
\draw[dash dot] (0,0) -- (0,7);
\draw[-] (0,0) -- (6,6) -- (3,9);
\draw[dash dot] (3,9) -- (3,7);
\draw[dashed] (3,7) -- (0,4);
\filldraw[black] (3,7) circle (1.5pt);
\draw[thick,black!80] (4.7,7.3) .. controls (4.2,7.8) and (2.5,6.5) .. (1.9,6);
\draw[very thick,blue] (5.75,6.25) -- (5.25,6.75);
\draw[very thick,blue] (5,7) -- (4.7,7.3);
\draw[very thick,black!80] (5,7) -- (5.25,6.75);
\draw[very thick,black!80] (5.75,6.25) -- (6,6);
\filldraw[blue] (5.75,6.25) circle (1.5pt);
\filldraw[blue] (5.25,6.75) circle (1.5pt);
\filldraw[blue] (5,7) circle (1.5pt);
\filldraw[blue] (4.7,7.3) circle (1.5pt);
\draw[very thick,red] (1,5.2) to[out=40,in=-145] (1.9,6);
\draw[very thick,black!80] (1,5.2) to[out=-140,in=0] (0,4.6);
\filldraw[red] (1,5.2) circle (1.5pt);
\filldraw[red] (1.9,6) circle (1.5pt);
\node at (3.5,3) {$\mathscr I^-$};
\node at (4.7,8) {$\mathscr I^+$};
\node[blue,rotate=-45] at (5.5,7.1) {$R$};
\node[red] at (1.4,6.1) {$I$};
\node at (3.4,6.4) {$\cal C$};
\end{tikzpicture}
\caption{\footnotesize The Penrose diagram for the evaporating black hole formed from collapse in an asymptotically flat spacetime. The Hawking radiation is collected in a subset $R\subset\mathscr I^+$ which lies in a Cauchy surface $\cal C$ that avoids the high curvature region near  the  singularity and the end-point of evaporation but includes any QES that can contribute. The functional integral is defined on the pink region from $\mathscr I^-$ up to the Cauchy surface where boundary conditions in the form of gluing conditions between copies of the spacetime are imposed in order to compute the (R\'enyi) entropies of the state on $R$.}
\label{fig1} 
\end{center}
\end{figure}
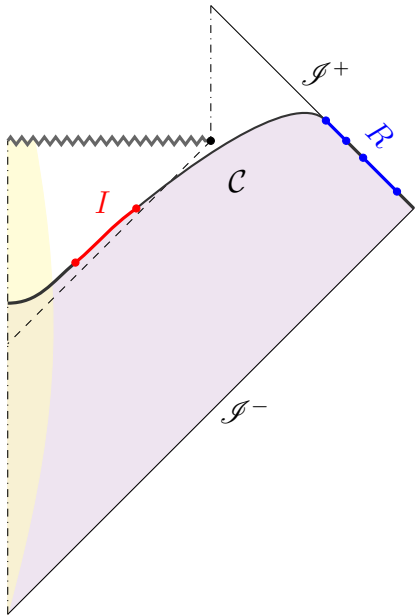
 
The calculations of \cite{Penington:2019kki,Almheiri:2019qdq} were done mainly in the context of the theory of Jackiw-Teitleboim (JT) gravity \cite{Jackiw:1984je,Teitelboim:1983ux} which captures the dominant $s$-wave sector of near-extremal Reissner-Nordstr\"om black holes in $3+1$. However, as advocated in \cite{Penington:2019npb} and then described in detail in \cite{Marolf:2020rpm}, conceptually the approach should apply to any black hole, like the Schwarzschild black hole in $3+1$ asymptotically flat spacetime. In that case the set up is illustrated in figure \ref{fig1} and $R$ is defined as a subset of $\mathscr I^+$. In the present work we will use the generalized entropy formalism as a hypothesis in the higher-dimensional case of an evaporating Schwarzschild black hole, along the lines of \cite{Marolf:2020rpm}. Of course, we emphasize that it can be fully justified in the context of JT gravity where the replica wormhole saddles can be found exactly \cite{Penington:2019kki,Almheiri:2019qdq}. We will present our results in such a way that they apply to both the JT gravity model and the Schwarzschild black hole. 

It has been argued \cite{Laddha:2020kvp, Raju:2020smc} that the holographic nature of gravity implies the fine grained entropy of the radiation is  actually a constant in asymptotically flat spacetimes and there is no Page curve (see also \cite{Geng:2020fxl, Geng:2021hlu,Chowdhury:2021nxw} for related issues). Our point of view is that within semiclassical gravity, suitably coarse-grained entropies or ``not-so-fine-grained" entropies as argued in \cite{Krishnan:2020oun, Ghosh:2021axl} would follow the usual Page curve \eqref{eq:page_theorem}. In this perspective, for the asymptotic observer the evaporation of a black hole is not too different from that of a piece of coal. In the case where AdS gravity  is coupled to a non-gravitating bath (as implied in the JT gravity setup) the graviton is not massless \cite{Geng:2020qvw} and the Page curve for the fine-grained entropy is crisply defined. 

The Lorentzian interpretation of the replica wormhole saddles is particularly interesting \cite{Marolf:2020rpm,Colin-Ellerin:2020mva}. It gives a picture where the entropy $S(R)$ includes the QFT entropy of an additional region $I$, an `entanglement island', behind the horizon. Since the modes inside and outside the horizon are entangled, as in the Unruh effect, this has a profound effect. The island is in the bulk where the metric can fluctuate and in the semi-classical limit the boundary of the island $\partial I$, the Quantum Extremal Surfaces  (QESs), is fixed dynamically by extremizing the `generalized entropy' \cite{Ryu:2006bv,Hubeny:2007xt,Faulkner:2013ana,Engelhardt:2014gca},\footnote{There is an important but subtle point that is worth emphasizing. Ordinarily, adding additional regions like the island $I$ to the entropy of a QFT would incur additional UV divergences from the QES which would suppress the contribution in the path integral. However, for the islands these divergences  are absorbed into the usual renormalization of Newton's constant in the QES terms in \eqref{guz2} so $S_I(R)$ has the same divergence as $S_\varnothing(R)$, the Hawking saddle. In this way, saddles with islands having multiple intervals can compete and sometimes dominate the entropy without being suppressed by a divergence.}
\EQ{
S_I(R)=\underset{\partial I}\ext\Big\{\sum_{\partial I}\frac{\text{Area}(\partial I)}{4G_N}+S_\text{QFT}(R\cup I)\Big\}
\ .
\label{guz2}
}
The QES are points in $1+1$ dimensions or $2d$ spatial surfaces in $3+1$. Finally, the entropy of $R$ is given by minimizing  over the possible island saddles:
\EQ{
S(R)=\min_{I} S_I(R)\ .
}
The island formula implies that there are underlying correlations in  the Hawking radiation that are not captured by the Hawking, i.e.~no island, saddle $S_\varnothing(R)\equiv S_\text{QFT}(R)$. A possible island $I$ is shown in figure \ref{fig1}, although strictly speaking only the QES, the endpoints, are covariantly defined and the island can be any Cauchy surface through the QES and containing $R$ on $\mathscr I^+$ like the one shown.

For most of the lifetime of a large black hole, the evaporation is very slow and there is an adiabatic limit for which it is possible to associate a slowly-varying temperature to the black hole. The adiabatic limit is valid when the Bekenstein-Hawking entropy of the black hole is suitably large,
\EQ{
\SBH(t)\gg{\cal N}\ ,
\label{eva}
}
where ${\cal N}$ is the number of fields (here free massless scalar fields) that make up the Hawking radiation. 
It is also assumed that ${\cal  N}\gg1$ to justify the semi-classical limit, an important ingredient in the replica wormhole saddle point analysis \cite{Penington:2019kki,Almheiri:2019qdq}. For the Schwarzschild black hole, the adiabatic approximation only breaks down near the end of the evaporation when the black hole becomes Planck sized% times $\sqrt{{\cal N}}$
. We will work in the limit where the intervals of radiation that make up $R\subset\mathscr I^+$, say $[u_j,u_{j+1}]$\footnote{The coordinate $u=t-r_*$ is the out-going null coordinate outside the black hole, where $r_*$ is the tortoise coordinate. Far from the black hole $r_*\approx r$. We choose the origin so that the black forms at $u=0$.}, are large compared with the thermal scale of the black hole $|u_j-u_{j+1}|\gg1/T$. In this limit, the QFT entropy is approximated by the thermodynamic entropy of a relativistic boson gas in $1+1$-dimensions with a slowly varying temperature,
\EQ{
S_\text{QFT}(R)\approx S_\text{rad}(R)+\text{(div)}\ ,
}
where $\text{(div)}$ is the usual UV divergence and where 
\EQ{
S_\text{rad}(u_j,u_{j+1})=\frac{\pi{\cal N}}6\int_{u_j}^{u_{j+1}}T\,du\ .
}
However, we will for the most part also work in a more stringent limit where the intervals are much larger than the scrambling time of the black hole $\Delta t_s$ which is the longer time scale
\EQ{
|u_j-u_{j+1}|\gg \Delta t_s\gg\frac1T\ ,\qquad \Delta t_s\thicksim\frac1T\log\frac{S_\text{BH}-S_*}{\cal N}\ .
\label{bxx}
}
Here, $S_*$ is the possible extremal entropy.

If we apply the adiabatic and the thermodynamic limit as above but combine it with the island formula then this leads to the very simple  islands-in-the-stream recipe for the saddles that can contribute \cite{Hollowood:2020kvk,Hollowood:2021nlo} to the case an evaporating black hole formed from a rapid collapse at $u=0$. In the JT gravity case, this involves a shockwave collapse into a pre-existing extremal black hole.
\begin{enumerate}[label=\protect\circled{\arabic*}]
\item Choose a subset $\partial\tilde I\subset\partial R\cup\{0^-\}$. Note that we explicitly add the point $u=0^-$ just before the black hole forms by a collapse at $u=0$.\footnote{This will account for the possible QES associated to the pre-existing extremal black hole in the case of JT gravity for which $S_\text{BH}(0^-)=S_*$, the extremal entropy, or for the Schwarzschild case, the origin of radial coordinate. In the latter case, the point is not a QES (so $\partial I$ there is not a boundary and $S_\text{BH}(0^-)=0$) but just accounts for the fact that island can extend smoothly over the origin of the polar coordinates.} The subset $\tilde I\subset\mathscr I^+$ is the `island-in-the-stream' that is the image of the island $I$ behind the horizon under the involution $U\to-U$ for the outgoing inertial Kruskal-Szekeres coordinate $U$, exchanging the inside and outside of the horizon.
\item The von Neumann entropy of $R$, up to a UV divergence independent of $I$, of this saddle is then
\EQ{
S_I(R)=\sum_{\partial\tilde I}\SBH (u_{\partial\tilde I})+S_\text{rad}(R\ominus\tilde  I)\ .
\label{ger}
}
Here, $\SBH(u)$ is the Bekenstein-Hawking entropy of the black hole. The set $R\ominus\tilde I$ is the symmetric difference of $R$ and $\tilde I$ on $\mathscr I^+$.
\end{enumerate}

The island-in-the-stream recipe above was derived under the simplifying assumption of trivial grey-body factor. 
One feature of a realistic black hole, as opposed to the simple model in JT gravity, is the fact that Hawking radiation must tunnel through an effective potential barrier around the black hole. This means that only some of the radiation makes it out to $\mathscr I^+$ and the Planck distribution gets modified from a black body to a grey body \cite{BD}
\EQ{
	\bar N_\omega^{\mathbb T}=\frac{\Gamma(\omega)}{e^{\omega/T}-1}\ ,
	\label{hrr}
}
where $\mathbb T$ indicates the modes transmitted through the potential barrier and $\bar N_\omega^{\mathbb T}$ is the expectation value of the occupation number of the mode with frequency $\omega$. The resulting grey-body factor $\Gamma(\omega)<1$ slows down the process of evaporation. 
For a black hole like Schwarzschild, the effect of the grey-body factor leads to simple multiplicative relation between the entropy rate of the transmitted radiation $S^{\mathbb T}_\text{rad}$\footnote{Here, $S_\text{rad}^{\mathbb T}$ is the QFT entropy of the original Hawking calculation after being suitably regularized and in the adiabatic limit.} and the one of the Bekenstein-Hawking entropy of the black hole
\EQ{
	\frac{dS^{\mathbb T}_\text{rad}}{du}=-\xi\frac{dS_\text{BH}}{du}\qquad \implies \qquad\frac{dS_\text{tot}}{du}=(\xi-1)\Big|\frac{dS_\text{BH}}{du}\Big|
	\label{rry}
}
where $1\leq\xi\leq2$ is the `grey-body coefficient'. Notice that $\xi\to1$ is the reversible limit, when only an infinitesimal amount of radiation escapes, while $\xi=2$ is the case with a trivial grey-body factor (no reflections). Since $\xi\geq1$, the rate of change of the total thermodynamic entropy is always $\geq0$ which is a statement of the generalized second law for black hole evaporation \cite{Bekenstein:1972tm,Bekenstein:1974ax}.

It is the purpose of this paper to analyse the effect of a non-trivial grey-body factor on entropy of an arbitrary subset of the radiation, and in particular how it affects the Page curve \cite{Page:1993wv,Page:2013dx} and the correlations of the Hawking radiation. Working in the adiabatic approximation \eqref{eva} and with intervals of radiation satisfying \eqref{bxx},\footnote{The condition of scales being large compared with the scrambling time can be relaxed, in which case the result \eqref{ger2} becomes more complicated.} the answer turns out to be a simple extension of the islands-in-the-stream formula \eqref{ger}:
\EQ{
\boxed{S_I(R)=\sum_{\partial\tilde I}\SBH(u_{\partial\tilde I})+S^{\mathbb T}_\text{rad}(R\ominus\tilde  I)\ .}
\label{ger2}
}
In the above, the entropy of the radiation is that of the radiation that tunnels out of the effective potential. So compared with \eqref{ger}, the second term involves the entropy of a relativistic bosonic gas with the modified Planck spectrum as in \eqref{hrr}. 

We can write a finite version of equation \eqref{rry}
\EQ{
S_\text{rad}^{\mathbb T}(u_1,u_2)=\xi\big(S_\text{BH}(u_1)-S_\text{BH}(u_2)\big)
}
and re-express \eqref{ger2} solely in terms of the Bekenstein-Hawking entropy:
\begin{equation}
	S_I(R)=\sum_{u_j\in \partial\tilde I} \SBH(u_j) + \xi \sum_{u_j\in \partial R \ominus \partial \tilde{I}} (-1)^{j+1} \SBH(u_j) \ ,
	\label{eq:gen3}
\end{equation}
with the $u_j$ ordered so that $u_j<u_{j+1}$.

If we take the view that the grey-body coefficient $\xi$ is a variable (e.g. by artificially reflecting back some of the radiation), then this allows us to investigate the effect of reversibility versus irreversibility as the evaporation proceeds. In particular, a key finding is that it is the irreversible nature of evaporation that leads to a picture where for an arbitrary subset of the radiation $R\subset\mathscr I^+$ there are multiple replica-wormhole saddles in \eqref{eq:gen3} that can compete to dominate the entropy, to compare with the reversible limit $\xi\to1$ where only one replica-wormhole saddle survives. This means that a na\"\i ve application of Page's theorem can account for the reversible limit but something rather more general is needed when the evaporation is irreversible $\xi>1$. Page's theorem can be proven by taking a unitary ensemble over the total state of the bipartite system and we will find that the generalization needed involves a nested sequence of unitary ensembles as the evaporation proceeds.

The organization of the paper is as fallows: in section \ref{s2}, we review some of the main features of Hawking radiation and the grey-body factor and how it influences the evaporation of the black hole. In particular, we describe how the entropy of the black hole and the radiation change as the black hole evaporates. In section \ref{s3}, we show how the entropy carried away by the radiation can be related to the entanglement of modes across the horizon. The analysis is complicated by the fact that when the grey-body factor is non-trivial there are modes that are reflected off the effective potential back into the black hole. A key feature is how the three sets of modes, the transmitted, the reflected and the entangled partner modes behind the horizon are all related. In section \ref{s4}, we consider the generalized entropy and island formalism when there is a grey-body factor in the adiabatic regime. This allows us to argue for the existence of a class of extrema of the generalized entropy and leads to a result that generalizes the analysis with a trivial grey-body factor in \cite{Hollowood:2021nlo}. The final sections are devoted to some applications of the formalism to topics that highlight the dependence on the grey-body factor. In section \ref{s5}, we argue that when the evaporation is irreversible, i.e.~$\xi>1$, the entropy of a subset of the Hawking radiation generally involves a competition of multiple saddles. We then describe how these multiple saddles can be understood by a nested generalization of Page's theorem. What results is a statistical quantum mechanical description of the evaporation. A key r\^ole is played by a time sequence of unitary averages. We uncover a concrete relation between the statistical model and the islands-in-the-stream formalism. In section \ref{s6}, we consider the effect of a grey-body factor on the entanglement-monogamy problem and highlight how the irreversible regime is rather richer than the reversible limit which manifests the so-called $A=R_B$ scenario.  Appendices \ref{a3} and \ref{a4} discuss the back-reaction problem of Hawking radiation on the geometry in the JT gravity and Schwarzschild cases, respectively. In the former case, the back-reaction problem can be solved exactly whereas for Schwarzschild there is an approximate analysis which is sufficient for our purposes.

\section{Grey-body factors and evaporation}\label{s2}

The mechanism that leads to Hawking radiation is a well understood textbook phenomenon \cite{BD}. It is  effectively a $1+1$ dimensional problem because each angular momentum mode can be treated separately and most of the emission energetically and entropically is in the $s$-wave mode. The JT  gravity model is formulated in $1+1$ dimensions \cite{Jackiw:1984je,Teitelboim:1983ux} and captures the dominant $s$-wave sector of the near-extremal Reissner-Nordstr\"om black hole in $3+1$ dimensions. The JT gravity model is useful because the problem of the back-reaction of Hawking radiation on the black hole, a crucial ingredient in modelling the evaporation of the black hole, can be solved exactly as we summarize in appendix \ref{a3}, even without invoking the adiabatic limit \cite{Engelsoy:2016xyb,Almheiri:2019psf,Hollowood:2020cou}. For the case of Schwarzschild, the back-reaction can be solved approximately by evoking the adiabatic limit \cite{Bardeen:1981zz,Parentani:1994ij,Massar:1994iy,Abdolrahimi:2016emo,Barcelo:2010xk} as we describe in appendix \ref{a4} which establishes some additional results that we need. What emerges from the analysis of an evaporating black hole in both JT gravity and Schwarzschild are some universal features which are precisely what we need in our analysis of the entropy of the radiation.

In the JT set up, the metric of the gravity region is a patch of AdS$_2$. A region of the boundary is then glued onto a half-Minkowski space \cite{Engelsoy:2016xyb,Almheiri:2019psf,Almheiri:2019qdq}. This is a model of the geometry of the $(t,r)$ part of the near-extremal Reissner-Nordstr\"om black hole in $3+1$ dimensions. So there are null coordinates $(u,v)$, $u=t-x$ and $v=t+x$ in the half-space Minkowski `bath' region $x>0$. These coordinates extend into the AdS region as the Schwarzschild coordinates that cover the region outside the horizon, where $x\equiv r_*$ is the tortoise coordinate. So $u\in[-\infty,\infty]$, where $u=\infty$ is the horizon. For the case of the Schwarzschild black hole, the coordinates $(u,v)$ are the analogues of Eddington-Finkelstein outgoing/ingoing coordinates that can be defined in the evaporating case. Far from the black hole, we have $u=t-r$ and $v=t+r$ since $r_* \sim r$.

In both JT gravity and Schwarzschild, we can also introduce the Kruskal-Szekeres (KS) type coordinates $(U,V)$ at least in a neighbourhood across the horizon. It is key observation that in this region these coordinates are related to $(u,v)$ via exponential maps
\EQ{
U=-e^{-\sigma(u)}\ ,\qquad V=e^{\sigma(v)}\  ,
\label{ruz}
}
where the function $\sigma(t)$ in the adiabatic limit is related to the slowly varying time-dependent temperature of the Hawking radiation via
\EQ{
\frac{d\sigma}{dt}=2\pi T\ ,
}
as discussed in the appendices.
Inside the horizon, we will introduce a null coordinate $\tilde u$, a partner to $u$ outside, with
\EQ{
U=e^{-\sigma(\tilde u)}\ .
}
There is an involution symmetry that exchanges the inside and the outside $U\leftrightarrow -U$, i.e. $u\leftrightarrow\tilde u$, that will play an important r\^ole in our analysis.

The metric in a neighbourhood of the horizon takes the form
\EQ{
ds^2\Big|_\text{near hor}=-\Omega_h(v)^{-2}dU\,dV\ ,
\label{kku}
}
where we have suppressed the angular part in the Schwarzschild case. The conformal factor is equal to $\Omega_h(v)=2\pi T(v)$ for the Schwarzschild case and constant for the JT gravity case.\footnote{Although for the $s$-wave reduction of the near-extremal charged black hole, there will be a similar factor to the Schwarzschild case.} The non-trivial conformal factor in the Schwarzschild case will be a subleading effect. The QES term in  \eqref{guz2} involves the area of an $S^2$ or, in the JT gravity case, the value of the dilaton. In both the Schwarzschild and JT gravity cases, in the neighbourhood of the horizon, i.e.~small $U$, we have the universal formula
\EQ{
\frac{\text{Area}(S^2)}{4G_N}\Big|_\text{near hor}=S_*+\big(S_\text{BH}(v)-S_*\big)(1-2UV)+\cdots\ .
\label{edd}
}
Here, $S_*$ is the entropy of the extremal black hole with $S_*=0$ in the Schwarzschild case.

In a black hole in $3+1$, modes that become Hawking radiation must tunnel out of an effective potential barrier that surrounds the black hole. The transmission probability for a mode of frequency $\omega$ is known as the grey-body factor $\Gamma(\omega)$. In general this depends on the angular momentum of the mode and the evaporation is dominated by the transmission of $s$-wave modes. In this work, we will exploit this to work in the $s$-wave approximation.

In the JT gravity set up, we can mimic the effect of the grey-body factor by choosing appropriate boundary conditions at the AdS-Minkowski interface so that modes are only partly transmitted. We can essentially engineer any grey-body factor we like. The modes of the scalar field satisfy the free wave equation and we can choose boundary conditions so that an incoming mode of frequency $\omega$ is partially transmitted at the boundary $x=0$:
\EQ{
\phi(x,t)=\begin{cases}e^{-i\omega u}+{\mathbb R}(\omega)e^{-i\omega v}\ , & x<0\ ,\\ {\mathbb T}(\omega)e^{-i\omega u}\ , & x>0\ .\end{cases}
\label{het2}
}
The grey-body factor is then the transmission probability
\EQ{
\Gamma(\omega)=|\mathbb T(\omega)|^2=1-|\mathbb R(\omega)|^2\ .
}
The phase of the transmission coefficient will lead to a time delay for the transmission of a wave packet.

The Hawking radiation carries away energy from the black hole which evaporates over time. The Hawking analysis is valid in the adiabatic limit when the evaporation is slow enough and  the black hole geometry can be described by a slowly time-dependent temperature. The time dependence is found by equating the rate of change of the mass to minus the outgoing energy flux of the radiation. The flux is determined by the occupation number \eqref{hrr}
\EQ{
\frac{dM}{du}=-{\cal N}\int_0^\infty\frac{d\omega}{2\pi}\cdot\frac{\omega\Gamma(\omega)}{e^{\omega/T}-1}\ .
\label{her}
}
The time dependence of the temperature then follows from the known relation between $T$ and $M$. The loss of mass also implies that the Bekenstein-Hawking entropy of the black hole $\SBH=\text{Area}(\text{horizon})/4G_N$, decreases in accordance the laws of thermodynamics
\EQ{
\frac{dS_\text{BH}}{du}=\frac1T\frac{dM}{du}\ .
\label{bur}
}
We can integrate this up from the endpoint of the evaporation $u=u_\text{evap}$  to get the entropy at an earlier time,
\EQ{
\SBH(u)\equiv\ZZ_u=S_*-\int^{u_\text{evap}}_u \frac1T\frac{dM}{du}\,du\ ,
\label{leg}
}
where $u_\text{evap}$ is the evaporation time which is infinite in the case of the charged black hole as it relaxes to the extremal black hole. 

In the adiabatic limit, the entropy of the radiation is given by the thermodynamic entropy of a relativistic bosonic gas in a `volume' $du$. The entropy density or flux is then
\EQ{
\frac{dS_\text{rad}}{du}={\cal N}\int_0^\infty\frac{d\omega}{2\pi}\big((\bar N_\omega+1)\log(\bar N_\omega+1)-\bar N_\omega\log\bar N_\omega\big)
\label{lod}
}
and so when there is a trivial grey-body factor $\Gamma(\omega)=1$, for an interval at $\mathscr I^+$ this gives
\EQ{
S_\text{rad}(u_1,u_2)=\frac{\pi{\cal N}}6\int_{u_1}^{u_2}T\,du\ .
}

Now we add in the grey-body factor. We will denote the entropy flux of the transmitted and reflected modes as $dS_\text{rad}^l/du$ where $l=\mathbb T$ or $\mathbb R$ which take the form \eqref{lod} with the appropriate occupation number:
\EQ{
\bar N_\omega^{\mathbb T}=\frac{\Gamma(\omega)}{e^{\omega/T}-1}\ ,\qquad \bar N_\omega^{\mathbb R}=\frac{1-\Gamma(\omega)}{e^{\omega/T}-1}\ .
\label{hrr2}
}

For the Schwarzschild black hole, there is only a single scale in the problem, the mass or the Schwarzschild radius $r_s=2G_NM$. The temperature takes the form
\EQ{
T=\frac1{8\pi G_NM}\ .
}
Because there is only one scale,  the grey-body factor  $\Gamma(\omega)$ is a function of the dimensionless ratio $\omega/T$, so we define
\EQ{
\tilde\Gamma(\omega/T)=\Gamma(\omega)\ .
}
Hence, we can write
\EQ{
\frac{dM}{du}=-\frac{\pi{\cal N}\eta T^2}{12}\ ,
}
where $\eta$ is just a number
\EQ{
\eta=\frac 6{\pi^2}\int_0^\infty dx\cdot\frac{x\tilde\Gamma(x)}{e^x-1}\ .
}
In addition, the entropy of the transmitted and reflected modes can be  expressed as
\EQ{
S^l_\text{rad}(u_1,u_2)=\frac{\pi{\cal N}\alpha^l}6\int_{u_1}^{u_2}T\,du \, , \qquad l = \mathbb{R,T} .
}
For instance,
\EQ{
\alpha^{\mathbb T}=\frac3{\pi^2}\int_0^\infty dx\,\left\{\Big(\frac{\tilde\Gamma(x)}{e^x-1}+1\Big)\log\Big(\frac{\tilde\Gamma(x)}{e^x-1}+1\Big)-\frac{\tilde\Gamma(x)}{e^x-1}\log\frac{\tilde\Gamma(x)}{e^x-1}\right\}
\label{fre}
}
and similarly for $\alpha^{\mathbb R}$ with  $\tilde\Gamma\to1-\tilde\Gamma$. Numerically for Schwarzschild in the $s$-wave approximation, 
\EQ{
\eta=0.86\ ,\qquad \alpha^{\mathbb T}=0.82\ ,\qquad \alpha^{\mathbb R}=0.31\ .
}
We will further define the grey-body coefficient $\xi$ as in \eqref{rry}, the ratio of the entropy fluxes of the black hole and the transmitted radiation, so $\xi=2\alpha^{\mathbb T}/\eta=1.91$.

If we allow for a more general grey-body factor, which is possible in the JT gravity model where $\Gamma(\omega)$ can be fixed by a choice of boundary condition between the AdS and Minkowski regions.
The upper bound $\xi=2$, i.e.~$\alpha^{\mathbb T}=\eta=1$ and $\alpha^{\mathbb R}=0$, corresponds to the case of a trivial grey-body factor $\Gamma(\omega)=1$. The lower bound $\xi=1$ can only be approached as a limit. For example, by taking the grey-body factor to switch on at high frequency $\Gamma(\omega)=\theta(\omega/T-\mu_0)$ for $\mu_0\gg1$, gives
\EQ{
\eta\approx\frac{6(\mu_0+1)}{\pi^2}e^{-\mu_0}\  ,\qquad\alpha^{\mathbb T}\approx\frac{3(\mu_0+2)}{\pi^2}e^{-\mu_0}\ ,
}
so that
\EQ{
\xi\approx1+\frac1{\mu_0}\ .
}
So in the limit of large $\mu_0$, the evaporation becomes very slow and the rate at which entropy is lost by the hole is equal to the entropy carried away in the radiation. In a thermodynamic sense this corresponds to a reversible situation for which $dS_\text{tot}/du\to0$, whilst if $\xi>1$ more entropy is carried away than is lost and the evaporation is thermodynamically irreversible. 

\section{Entropy and entanglement}\label{s3}

In this section we turn to the entropy dynamics in the evaporating black hole background.

\subsection{Entanglement across the horizon}

The entanglement of modes across the horizon is the physics of the Unruh effect. The outgoing modes are in the $U$ vacuum, the inertial frame across the horizon, which can be expressed in terms of the $u$ vacuum, the one appropriate to the asymptotic region far from the hole, as a 2-mode squeezed state. Schematically,
\EQ{
\ket{0}_U\thicksim\prod_{\omega} \exp\Big[e^{-\omega/2T}a^\dagger({\cal Z}_{\omega})a^\dagger(\widetilde {\cal Z}_{\omega}^*)\Big]\ket{0}_u\ .
\label{bot}
}
In this expression, ${\cal Z}_{\omega}$ are a complete set  of modes outside the horizon and $\widetilde {\cal Z}_{\omega}$ are their partners behind the horizon related by the involution $U\leftrightarrow-U$ that exchanges the inside of the horizon with the outside.

If we trace out the modes behind horizon, the state of outgoing modes that reach $\mathscr I^+$ is a thermal state of temperature $T$:
\EQ{
\rho={\EuScript N}\exp\Big[-\frac 1T\sum_\omega\omega a^\dagger({\cal Z}_{\omega})a({\cal Z}_{\omega})\Big]=
{\EuScript N}\prod_\omega\sum_{n=0}^\infty e^{-n\omega/T}\ket{n}\bra{n}\ ,
\label{hit}
}
where ${\EuScript N}=\prod_\omega(1-e^{-\omega/T})$. So the probability of occupation of the $n^\text{th}$ level  of a mode of frequency $\omega$ is
\EQ{
p_n=(1-e^{-\omega/T})e^{-n\omega/T}\ ,
\label{poo}
}
implying that the mean occupation number is
\EQ{
\bar N_\omega=\sum_{n=0}^\infty np_n=\frac1{e^{\omega/T}-1}\ .
\label{pun}
}

In the Unruh state \eqref{bot}, we can split the modes outside the horizon into a sum of a reflected and transmitted component, 
\EQ{
a({\cal Z}_\omega)=a({\cal Z}^{\mathbb R}_{\omega})+a({\cal Z}^{\mathbb T}_{\omega})\ .
\label{bo2}
}
We now have a tripartite systems of modes and we can calculate the state of any subset by tracing out the other two. For example, if we trace out the modes behind the horizon $\widetilde{\cal Z}$ and the reflected modes ${\cal Z}^{\mathbb R}$ then we find the state of a transmitted mode of frequency $\omega$ as in \eqref{hit} but with a  probability,
\EQ{
p_n^{\mathbb T}=\frac{(1-e^{-\omega/T})e^{-n\omega/T}\Gamma(\omega)^n}{(1-e^{-\omega/T}(1-\Gamma(\omega)))^{n+1}}\ ,
\label{nun}
}
where the grey-body factor arises as the normalization of the transmitted mode. So the state of an individual mode is thermal but with a temperature that is no longer $T$ since it depends on the grey-body factor $\Gamma(\omega)$. This probability gives a mean occupation number modified by the grey-body factor precisely as in \eqref{hrr}. 

We can also trace out the modes behind the horizon and the transmitted modes to give the occupation number  of the reflected modes as in \eqref{hrr2}. Finally, if we trace out the reflected and transmitted modes, then the occupation number of the modes behind the horizon is simply $\bar N_\omega=1/(e^{\omega/T}-1)$.

For an evaporating black hole, we will assume that the infalling modes are in the vacuum of the inertial coordinate $v$ far from the black hole.

\subsection{Entropy}

Let us analyse the entropy of intervals either outside or inside the horizon. To start with we ignore the grey-body factor, so there is no reflection.

\paragraph{Trivial grey-body factor:} the entropy of an interval $R=[p_1,p_2]$ outside the horizon far from the black hole is given by \cite{Calabrese:2004eu}
\EQ{
S_\text{QFT}(R)=\frac{\cal N}6\log(U_2-U_1)(v_1-v_2)-\frac{\cal N}{12}\sum_{i=1}^2\log(2\pi T(u_i)U_i)+\text{(div)}\ .
}
The second term here are the conformal factors for the conformal transformation $u\to U$ at the endpoints of $R$. In the adiabatic limit, we can discard the term involving $T$ and the infalling mode contribution involving $v_2-v_1$ since these are subleading. This is  to be expected because the infalling modes are in the $v$ vacuum. 

When the interval is sufficiently large $|U_1|\gg|U_2|$
\EQ{
S_\text{QFT}(R)\approx\frac{\cal N}6\log\sqrt{\frac{U_1}{U_2}}+\text{(div)}=\frac{\pi{\cal N}}6\int^{u_2}_{u_1}T\,du+\text{(div)}\equiv S_\text{rad}(R)+\text{(div)}\ .
\label{kso}
}
So in the limit we are working, the QFT entropy is approximately equal to the thermodynamic entropy of the radiation, up to the usual UV divergence.

Now we turn to an interval behind the horizon (an island) $I=[p_1,p_2]$ that is specifically in the near-horizon zone $|UV|\ll1$, so that the metric is approximately given in \eqref{kku}. The entropy can be calculated using standard QFT techniques given that the quantum state is the Unruh vacuum:
\EQ{
S_\text{QFT}(I)=\frac{\cal N}6\log(U_2-U_1)(v_1-v_2)+\frac{\cal N}{12}\log\frac{(2\pi)^2 T(v_1)V_1T(v_2)V_2}{\Omega_h(v_1)\Omega_h(v_2)}+\text{(div)}\ .
}
In the generalized entropy application, the UV divergence here will be absorbed into the usual one-loop renormalization of Newton's constant. The terms $\log(v_1-v_2)$ and $\log(2\pi T/\Omega_h)$ are subleading in the adiabatic limit and can be dropped. In the thermodynamic limit, we can write this as
\EQ{
S_\text{QFT}(I)\approx\tilde S(I)+S_\text{rad}(I)+\text{(div)}\approx S_\text{rad}(I)+\text{(div)}\ ,
\label{lee}
}
where we define
\EQ{
\tilde S(I)=\frac{\cal N}{12}\sum_{\partial I}\log(U_{\partial I}V_{\partial I})\ ,
\label{ugb}
}
a term that will also turn out to be subleading in the adiabatic/thermodynamic limit but plays an important r\^ole in the extremum conditions for the generalized entropy. We conclude that the entropy of an interval either inside \eqref{lee} or outside \eqref{kso} the horizon is just the thermodynamic entropy of the outgoing modes that cross the interval.

\begin{figure}[ht]
\begin{center}
\begin{tikzpicture} [scale=0.9]
\filldraw[blue!5] (2.5,4.5) -- (5,7) -- (4.5,7.5) -- (2.5,5.5) -- cycle;
\filldraw[blue!20] (2,4) -- (2.5,4.5) -- (2.5,5.5) -- (1,4) -- cycle;
\filldraw[blue!20] (2,5) -- (2.5,5.5) -- (1,7) -- (0,7) -- cycle;
\draw[decorate,very thick,black!60,decoration={zigzag,segment length=1.5mm,amplitude=0.5mm}] (0,7) -- (3,7);
\draw[dash dot] (0,4) -- (0,7);
\draw[-] (4,4) -- (6,6) -- (3,9);
\draw[dash dot] (3,9) -- (3,7);
\draw[dashed] (3,7) -- (0,4);
\filldraw[black] (3,7) circle (1.5pt);
\draw[very thick,blue] (5,7) -- (4.5,7.5);
\filldraw[blue] (5,7) circle (1.5pt);
\filldraw[blue] (4.5,7.5) circle (1.5pt);
\draw[very thick,blue] (1.2,5.8) to[out=20,in=-145] (1.9,6.1);
\filldraw[blue] (1.2,5.8) circle (1.5pt);
\filldraw[blue] (1.9,6.1) circle (1.5pt);
\node[blue,rotate=0] at (5.1,7.6) {$\hat I$};
\node[blue] at (1.4,6.2) {$I$};
\begin{scope}[xshift=-8cm]
\filldraw[red!5] (2.5,4.5) -- (5,7) -- (4.5,7.5) -- (2.5,5.5) -- cycle;
\filldraw[red!5] (2,4) -- (2.5,4.5) -- (2.5,5.5) -- (1,4) -- cycle;
\filldraw[red!20] (0.7,5.3) -- (2.4,7) -- (2.8,7) -- (0.7,4.9) -- cycle;
\draw[decorate,very thick,black!60,decoration={zigzag,segment length=1.5mm,amplitude=0.5mm}] (0,7) -- (3,7);
\draw[dash dot] (0,4) -- (0,7);
\draw[-] (4,4) -- (6,6) -- (3,9);
\draw[dash dot] (3,9) -- (3,7);
\draw[dashed] (3,7) -- (0,4);
\filldraw[black] (3,7) circle (1.5pt);
\draw[very thick,red] (5,7) -- (4.5,7.5);
\filldraw[red] (5,7) circle (1.5pt);
\filldraw[red] (4.5,7.5) circle (1.5pt);
\draw[very thick,red] (1.2,5.8) to[out=20,in=-145] (1.9,6.1);
\filldraw[red] (1.2,5.8) circle (1.5pt);
\filldraw[red] (1.9,6.1) circle (1.5pt);
\node[red,rotate=0] at (5,7.5) {$\tilde I$};
\node[red] at (1.4,6.3) {$I$};
\end{scope}
\end{tikzpicture}
\caption{\footnotesize On the left the outgoing modes inside the horizon that pass through the island and their map onto $\tilde I\subset\mathscr I^+$ via the involution $\tilde u\to u$ or $U\to-U$. On the right, we illustrate the reflected modes that pass through the island $I$ and their map onto $\hat I\subset\mathscr I^+$ via $v\to u$. }
\label{fig2} 
\end{center}
\end{figure}
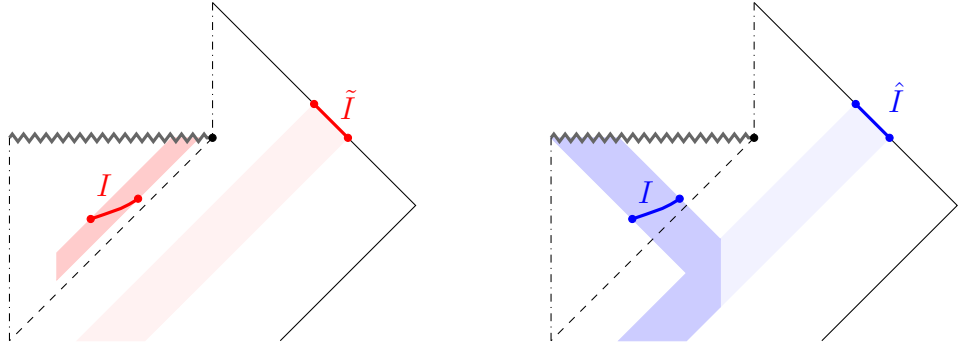

Things get more interesting when we consider the entropy of intervals both inside and outside the horizon $S_\text{QFT}(R\cup I)$. For intervals that are large compared with the thermal scale, $\Delta u\gg T^{-1}$, the entanglement between modes inside and outside the horizon is approximately local in null space and so may be determined by a process of ray tracing. So an interval of outgoing modes inside the horizon $I=[\tilde u_1,\tilde u_2]$ is purified with its mirror interval $\tilde I=[u_1,u_2]\subset\mathscr I^+$, with $\tilde u_i=u_i$: see figure \ref{fig2}. Hence, modes in $R\cap\tilde I\subset\mathscr I^+$ are purified by modes behind the horizon in $I$ and so do not contribute to the entropy. It follows that the net entropy corresponds to the modes in the symmetric difference:
\EQ{
S_\text{QFT}(R\cup I)\approx S_\text{rad}(R\ominus\tilde I)+\text{(div)}\ ,
\label{ewe}
}
where $R\ominus\tilde I=R\cup\tilde I-R\cap\tilde I$.
 
\paragraph{With grey-body factor:}now there is non-trivial reflection and the modes collected in the interval $R$ far from the black hole are the modes transmitted by the potential barrier and, hence, \eqref{kso} is replaced by
\EQ{
S_\text{QFT}(R)\approx\int_{u_1}^{u_2}\frac{dS_\text{rad}^{\mathbb T}}{du}\,du\equiv S^{\mathbb T}_\text{rad}(R)+\text{(div)}\ ,
}
where the entropy flux or density for the transmitted and reflected modes is defined in \eqref{lod} with occupation number $\bar N_\omega^{\mathbb T}$.

The island, being behind the horizon, now also picks up modes that are reflected off the potential barrier at $u=v$. The effect of these modes on $S_\text{QFT}(R\cup I)$ can be considered after mapping the reflected modes that pass through $I$ onto $\mathscr I^+$ by using the map $v\to u$. So if the $v$ coordinate of $I$ are $[v_1,v_2]$ then we define $\hat I=[v_1,v_2]\subset\mathscr I^+$: see figure \ref{fig2}. 

\begin{figure}[ht]
\begin{center}
\begin{tikzpicture} [scale=0.8]
\draw[fill=blue!20] (1.732,-1) circle (1.5);
\draw[fill=red!20] (1.732,1) circle (1.5);
\draw[fill=green!20] (0,0) circle (1.5);
\begin{scope}
\clip (1.732,1) circle (1.5);
\fill[blue!20] (0,0) circle (1.5);
\end{scope}
\begin{scope}
\clip (1.732,-1) circle (1.5);
\fill[red!20] (0,0) circle (1.5);
\end{scope}
\begin{scope}
\clip (1.732,-1) circle (1.5);
\fill[green!20] (1.732,1) circle (1.5);
\end{scope}
\begin{scope}
\clip (1.732,1) circle (1.5);
\clip (1.732,-1) circle (1.5);
\fill[white] (0,0) circle (1.5);
\end{scope}
\draw (1.732,-1) circle (1.5);
\draw (1.732,1) circle (1.5);
\draw (0,0) circle (1.5);
\node at (-0.2,0) {\Large $R$};
\node at (1.9,1.1) {\Large $\tilde I$};
\node at (1.9,-1.1) {\Large $\hat I$};
\begin{scope}[xshift=4.5cm]
\node[right,draw=black,fill=red!20,rounded corners=3pt] at (0,1.5) {$S_\text{rad}$};
\node[right,draw=black,fill=green!20,rounded corners=3pt] at (0,0) {$S_\text{rad}^{\mathbb T}$};
\node[right,draw=black,fill=blue!20,rounded corners=3pt] at (0,-1.5) {$S_\text{rad}^{\mathbb R}$};
\end{scope}
\end{tikzpicture}
\caption{\footnotesize The contributions to $S_\text{QFT}(R\cup I)$ as determined by the overlaps of the three subsets of $\mathscr I^+$ labelled $R$, $\tilde I$ and $\hat I$. Note that the triple intersection does not contribute because the modes in the 3 subsets together are in a pure state.}
\label{fig3} 
\end{center}
\end{figure}
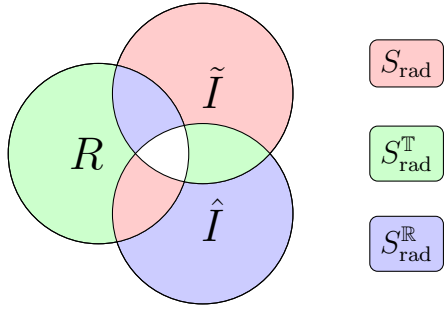

The contribution to $S_\text{QFT}(R\cup I)$ in the thermodynamic limit now depends on the interplay of the three subsets $R$, $\tilde I$ and $\hat I$ of $\mathscr I^+$. The contributions depends on the tripartite entanglement structure that is summarized in figure \ref{fig3}. Explicitly,
 \EQ{
S_\text{QFT}(R\cup I)&\approx S_\text{rad}((\tilde I\ominus R)\cap(\tilde I\ominus\hat I))+S^{\mathbb T}_\text{rad}((R\ominus\tilde I)\cap(R\ominus\hat I))\\ &+S^{\mathbb R}_\text{rad}((\hat I\ominus R)\cap(\hat I\ominus\tilde I))+\text{(div)}\ ,
\label{pew}
}
generalizing \eqref{ewe}\footnote{It would be interesting to derive such a result from purely CFT arguments following the analysis in \cite{Kruthoff:2021vgv}, where the entanglement entropy of an interval containing the semi-reflecting barrier was computed in some limits.}.
 
However, if we work in the limit that the intervals are large compared with the scrambling time \eqref{bxx}, we will find in section \ref{s4} that relevant configurations have islands $I$ with endpoints, the QES, that have outgoing and infalling coordinates related at leading order by\footnote{There are important subleading corrections to the following which will not affect  the entropy at leading order we are working.}
\EQ{
\tilde u=v\ .
}
The implication for the entropy is that $\tilde I=\hat I$ at leading order and so \eqref{pew} simplifies to
\EQ{
S_\text{QFT}(R\cup I)\approx S^{\mathbb T}_\text{rad}(R\ominus\tilde I)+\text{(div)}\ .
\label{ewe2}
}

\section{Islands in the adiabatic limit}\label{s4}

In this section, we prove that refinement of the `islands in the stream' formula for the generalized entropy \eqref{ger2}. Specifically we prove the following for the coordinates $(\tilde u,v)$ of the QES:
\begin{enumerate}[label=\protect\circled{\arabic*}]
\item The $\tilde u$ coordinate is equal to the $u$ coordinates of one of the endpoints $\partial R$ at $\mathscr I^+$. More precisely
\EQ{
\tilde u=u_j+{\cal O}(T^{-1})\ ,\qquad u_j\in\partial R\ .
}
The one exception is that there can be QES just before the black hole is formed. In the JT gravity case, this is the QES of the extremal black hole, while for Schwarzschild it is just the origin of the polar coordinates .
\item The infalling and outgoing coordinates of a QES are related by
\EQ{
v=\tilde u-\varepsilon\ ,
\label{ler}
}
where $\varepsilon={\cal O}(T^{-1}\log s^{-1})$ is a subleading correction of order the scrambling time \eqref{bxx} and we have defined 
\EQ{
s\equiv\frac{\cal N}{\SBH(u)-S_*}\ll1\ .
}
Note that in the limit we are working the intervals are always large \eqref{bxx} and so $\varepsilon$ although large is subleading compared to the size of the intervals $R$. The temperature varies very slowly and so the combination $UV$ for a QES is very small since
\EQ{
UV=e^{-\sigma(\tilde u)+\sigma(\tilde u-\varepsilon)}\approx e^{-2\pi T(\tilde u)\varepsilon}\approx s\ll1\ .
\label{bix}
}
It follows that the QES lie close to the horizon. 
\end{enumerate}

The entropy of this saddle can then be evaluated as follows at leading order in the adiabatic limit. The first term in the generalized entropy, the expression in the curly brackets in \eqref{guz2},  is just the Bekenstein-Hawking entropy of the black hole evaluated at the infalling coordinate $v_{\partial I}$, which we can relate to the outgoing coordinate of $\partial\tilde I$ \eqref{ler}, that is $\SBH(v_{\partial I})\approx\SBH(u_{\partial\tilde I})$. The second term is the QFT entropy of $R\cup I$. If the QES takes the form \eqref{ler}, then we get that $S_\text{QFT}(R\cup I)$ is given by \eqref{ewe2}, which in turns leads to the main result \eqref{ger2}.

\subsection{Proving extremality}

In order to establish our conjecture, namely the pattern of the extrema and the leading order expression for the entropy \eqref{ger}, we need to show that variations of the generalized entropy around the putative extrema vanish.

Let we label the points in $\partial R$ with outgoing coordinates $u_j$, $j=1,2\ldots, n$, ordered so that
\EQ{
u_1<u_2<\cdots<u_n\ ,
}
so that $R\subset\mathscr I^+$ consists of the intervals $[u_1,u_2]\cup[u_3,u_4]\cup\cdots\cup[u_{n-1},u_n]$. Let us label the QES $\partial I$ with coordinates $(U_a,V_a)$ or $(\tilde u_a,v_a)$, $a= 1, \dots , p$.

The fact that $\partial\tilde I\subset\partial R$, corresponds to a one-to-one map of the QES to the endpoints  of $R$ defined by $\alpha(a)$ such that
\EQ{
\tilde u_a=u_{\alpha(a)}\ ,\qquad v_a=u_{\alpha(a)}\ .
\label{weq}
}

Now let us concentrate on the $a^\text{th}$ QES and shift around the leading order expressions \eqref{weq}
\EQ{
\tilde u_a=u_{\alpha(a)}+\delta\tilde u_a\ ,\qquad v_a=u_{\alpha(a)}+\delta v_a\ .
}
The shifts are not infinitesimal but are small compared with the size of the intervals. The shift $\delta v_a$ is in the infalling coordinate of the QES and we will see that it is negative, $\delta v_a = - \Delta t_s$, where $\Delta t_s$ is  interpreted as the instantaneous scrambling time of the black hole as in \eqref{bxx}.

Our goal will be to show that there exists a solution for $\delta\tilde u_a$ and $\delta v_a$ but the actual values will not be needed because they are subleading in the limit we are working with. Recalling that $U_{\alpha(a)}<0$, we will write,
\EQ{
U_a=-\lambda_a U_{\alpha(a)}\ ,\qquad V_a=-\mu_a/U_{\alpha(a)}\ ,
}
where
\EQ{
\lambda_a=e^{2\pi T\delta\tilde u_a}\ ,\qquad \mu_a=e^{2\pi T\delta v_a}\ .
}
Here, the slowly varying function $T$ is implicitly evaluated at $u_{\alpha(a)}$. 

We will make the hypothesis that at the extremum
\EQ{
\mu_a=\mathscr O(s^1)\ ,\qquad\lambda_a=\mathscr O(s^0)
\label{cyr}
}
and show that a solution exists which is consistent with this assumption. The implication is that
$U_aV_a=\lambda_a\mu_a=\mathscr O(s)\ll1$ which means that the QES is very close to the horizon on the inside. The behaviour of $\mu_a$ implies that the scrambling time
\EQ{
\Delta t_s\thicksim\frac1{2 \pi T(u_{\alpha(a)})}\log \frac{\SBH(u_{\alpha(a)})-S_*}{\cal N}\ .
}

There are two contributions to the entropy gradients of the generalized entropy \eqref{guz2}. Firstly, the QES term. Since the QES are, by hypothesis, close to the horizon $U_aV_a\ll1$, we can use the near-horizon approximation \eqref{edd} for the first term in \eqref{guz2}:
\EQ{
\sum_{\partial I}\frac{\text{Area}(\partial I)}{4G_N}=pS_*+\sum_a(\SBH(v_a)-S_*)(1-2U_aV_a)+\cdots\ .
\label{pul}
}
This expression is universal and so valid for the near-extremal RN and Schwarzschild black holes.  Using this, the gradient in the direction $\tilde u_a$ is
\EQ{
\frac\partial{\partial\tilde u_a}\sum_{\partial I}\frac{\text{Area}(\partial I)}{4G_N}=\frac{4\pi T\lambda_a\mu_a}s+\cdots\ .
\label{voo}
}
On the other hand, the gradient in the direction $v_a$ is
\EQ{
\frac\partial{\partial v_a}\sum_{\partial I}\frac{\text{Area}(\partial I)}{4G_N}=-\frac{4\pi T \lambda_a\mu_a}s+\frac{dS_\text{BH}(v_a)}{dv_a}+\cdots\ ,
\label{woo}
}
Note that to leading order we can replace $\tilde u_a$ and $v_a$ with $u_{\alpha(a)}$ in the arguments of $\SBH$ and $T$ since in the adiabatic limit these are slowly varying. In the following, all the slowly varying functions $T$, $\dot S_\text{BH}=dS_\text{BH}/du$ and $\dot S_\text{rad}=dS_\text{rad}/du$, etc, without an argument are implicitly evaluated at $u_{\alpha(a)}$. 

The second contribution comes from the variation of the QFT entropy term $S_\text{QFT}(R\cup I)$. Let us consider its variation under $\delta\tilde u_a$ and visualize the effect on the modes collected by $R\cup I$ that we map to $\mathscr I^+$ as $R$, $\hat I$ and $\tilde I$:
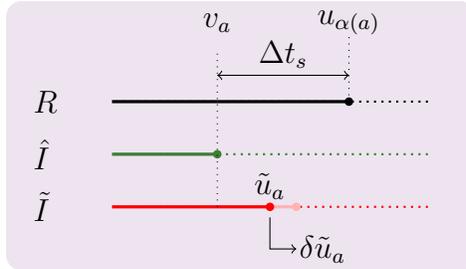
\begin{center}
\begin{tikzpicture} [scale=0.7]
\filldraw[fill = Plum!10!white, draw = Plum!10!white, rounded corners = 0.2cm] (-2,1.9) rectangle (6.9,-3.2);
\begin{scope}[xscale=-1,xshift=-6cm]
\draw[dotted,thick] (0,0) -- (6,0);
\draw[dotted,thick,OliveGreen] (0,-1) -- (6,-1);
\draw[dotted,thick,red] (0,-2) -- (6,-2);
\draw[very thick] (1.5,0) -- (6,0);
\draw[OliveGreen,very thick] (4,-1) -- (6,-1);
\draw[very thick,red] (3,-2) -- (6,-2);
\draw[very thick,red!30] (2.5,-2) -- (3,-2);
\filldraw[black] (1.5,0) circle (2pt);
\filldraw[OliveGreen] (4,-1) circle (2pt);
\filldraw[red!30] (2.5,-2) circle (2pt);
\filldraw[red] (3,-2) circle (2pt);
\draw[->] (3,-2.2) -- (3,-2.8) -- (2.5,-2.8);
\node at (2,-2.8) {$\delta\tilde u_a$};
\node at (3,-1.6) {$\tilde u_a$};
\draw[dotted] (4,-2) -- (4,1);
\draw[dotted] (1.5,0) -- (1.5,1.3);
\draw[<->] (1.5,0.5) -- (4,0.5);
\node at (1.5,1.5) {$u_{\alpha(a)}$};
\node at (4,1.5) {$v_a$};
\node at (2.75,0.9) {$\Delta t_s$};
\end{scope}
\node[right] at (-1.7,0) {$R$};
\node[right] at (-1.7,-1) {$\hat I$};
\node[right] at (-1.7,-2) {$\tilde I$};
\end{tikzpicture}
\end{center}
The variation of the QES by its $\tilde u$ coordinates affects $\tilde I$ but fixes $\hat I$. It is apparent that the variation converts modes in $R$ into modes in $R\cup\tilde I$, i.e.~leading to a gradient $\dot S_\text{rad}^{\mathbb R}-\dot S_\text{rad}^{\mathbb T}$. Hence, 
\EQ{
\frac{\partial S_\text{QFT}(R\cup I)}{\partial \tilde u_a}=\frac{\partial\tilde S(I)}{\partial\tilde u_a}-\dot S_\text{rad}^{\mathbb T}+\dot S_\text{rad}^{\mathbb R}=-\frac{\pi cT}6-\dot S_\text{rad}^{\mathbb T}+\dot S_\text{rad}^{\mathbb R}\ .
\label{moo}
}
The first term here comes from the conformal factor defined in \eqref{ugb}. Although the value of this term is subleading at the saddle point, its gradient cannot be ignored. Using the expression \eqref{lod}, we can write this term as the flux $\dot S_\text{rad}$, i.e.~with no grey-body factor, so finally
\EQ{
\frac{\partial S_\text{QFT}(R\cup I)}{\partial \tilde u_a}=-\dot S_\text{rad}-\dot S_\text{rad}^{\mathbb T}+\dot S_\text{rad}^{\mathbb R}\qquad (\tilde u_a<u_{\alpha(a)})\ .
\label{moo2}
}

Now we can follow the same reasoning for the case when $\tilde u_a>u_{\alpha(a)}$:
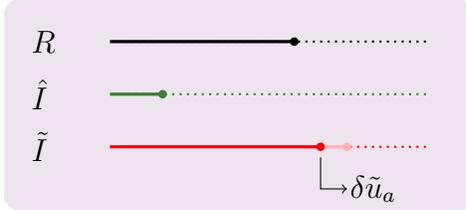
\begin{center}
\begin{tikzpicture} [scale=0.7]
\filldraw[fill = Plum!10!white, draw = Plum!10!white, rounded corners = 0.2cm] (-2,0.8) rectangle (6.9,-3.2);
\begin{scope}[xscale=-1,xshift=-6cm]
\draw[dotted,thick] (0,0) -- (6,0);
\draw[dotted,thick,OliveGreen] (0,-1) -- (6,-1);
\draw[dotted,thick,red] (0,-2) -- (6,-2);
\draw[very thick] (2.5,0) -- (6,0);
\draw[very thick,OliveGreen] (5,-1) -- (6,-1);
\draw[very thick,red] (2,-2) -- (6,-2);
\draw[very thick,red!30] (1.5,-2) -- (2,-2);
\filldraw[black] (2.5,0) circle (2pt);
\filldraw[OliveGreen] (5,-1) circle (2pt);
\filldraw[red!30] (1.5,-2) circle (2pt);
\filldraw[red] (2,-2) circle (2pt);
\draw[->] (2,-2.2) -- (2,-2.8) -- (1.5,-2.8);
\node at (1,-2.8) {$\delta\tilde u_a$};
\end{scope}
\node[right] at (-1.7,0) {$R$};
\node[right] at (-1.7,-1) {$\hat I$};
\node[right] at (-1.7,-2) {$\tilde I$};
\begin{scope}[xshift=8cm,yshift=-1.8cm]
%\node[right,fill=Plum!10!white,draw = Plum!10!white,rounded corners=3pt] at (0,1) {$\delta\tilde I>0$};
\end{scope}
\end{tikzpicture}
\end{center}
\noindent In this case, the variation is
\EQ{
\frac{\partial S_\text{QFT}(R\cup I)}{\partial \tilde u_a}=\frac{\partial\tilde S(I)}{\partial\tilde u_a}+\dot S_\text{rad}=0\qquad (\tilde u_a>u_{\alpha(a)})\ .
\label{noo}
}

Our analysis above seems to imply a discontinuous change in the gradient at $\tilde u_a=u_{\alpha(a)}$. This is simply a symptom of the fact that we have used the thermodynamic approximation for the entropies which will not be valid when $\tilde u_a$ and $u_{\alpha(a)}$ are close. Fortunately, we do  not need to have an explicit expression for the smooth crossover in order to establish the existence of extrema and so it will be sufficient for our purposes to write 
\EQ{
\frac{\partial S_\text{QFT}(R\cup I)}{\partial \tilde u_a}=\big(\dot S_\text{rad}^{\mathbb R}-\dot S_\text{rad}^{\mathbb T}-\dot S_\text{rad}\big)\Theta(\lambda_a-1)\ .
\label{noo2}
}
for some smooth function $\Theta(x)$ that goes from 0 to 1 as $x\in[-1,\infty]$ goes from negative to positive values, i.e.~a smoothed version of the Heaviside function. 

As a useful consistency check, in the case with a trivial grey-body factor, where there are no reflected modes, we can calculate the entropy gradient exactly using exact QFT results based on conformal transformations \cite{Hollowood:2021nlo}:
\EQ{
\frac{\partial S_\text{QFT}(R\cup I)}{\partial \tilde u_a}=-\frac{\pi \mathcal{N} T}3\cdot\frac{U_a}{U_a-U_{\alpha(a)}}=-2\dot S_\text{rad}\cdot\frac{\lambda_a}{\lambda_a+1}\ ,
}
which matches the behaviour \eqref{noo2} when $S_\text{rad}^{\mathbb R}=0$ and $S_\text{rad}^{\mathbb T}\equiv S_\text{rad}$ if we identify
\EQ{
\Theta(\lambda_a-1)=\frac{\lambda_a}{\lambda_a+1}\ .
}
In the presence of a grey-body factor we cannot use conformal transformations in the same way to calculate the cross over exactly.

Now consider the variation $\delta v_a$. In this case there is only one configuration:
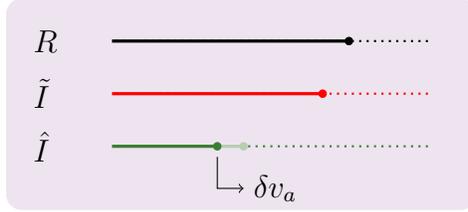
\begin{center}
\begin{tikzpicture} [scale=0.7]
\filldraw[fill = Plum!10!white, draw = Plum!10!white, rounded corners = 0.2cm] (-2,0.8) rectangle (6.9,-3.2);
\begin{scope}[xscale=-1,xshift=-6cm]
\draw[dotted,thick] (0,0) -- (6,0);
\draw[dotted,thick,OliveGreen] (0,-2) -- (6,-2);
\draw[dotted,thick,red] (0,-1) -- (6,-1);
\draw[very thick] (1.5,0) -- (6,0);
\draw[very thick,OliveGreen] (4,-2) -- (6,-2);
\draw[very thick,red] (2,-1) -- (6,-1);
\draw[very thick,OliveGreen!30] (3.5,-2) -- (4,-2);
\filldraw[black] (1.5,0) circle (2pt);
\filldraw[OliveGreen] (4,-2) circle (2pt);
\filldraw[OliveGreen!30] (3.5,-2) circle (2pt);
\filldraw[red] (2,-1) circle (2pt);
\draw[->] (4,-2.2) -- (4,-2.8) -- (3.5,-2.8);
\node at (2.9,-2.8) {$\delta v_a$};
\end{scope}
\node[right] at (-1.7,0) {$R$};
\node[right] at (-1.7,-2) {$\hat I$};
\node[right] at (-1.7,-1) {$\tilde I$};
%
%
%\begin{scope}[xshift=8cm,yshift=-1.5cm]
%\node[right,fill=Plum!10!white,draw = Plum!10!white,rounded corners=3pt] at (0,1) {$\delta(R\cup\tilde I)>0$};
%\node[right,fill=Plum!10!white,draw = Plum!10!white,rounded corners=3pt] at (0,0) {$\delta(R^{\mathbb T}\cup\tilde I)<0$};
%\end{scope}
\end{tikzpicture}
\end{center}
\noindent The variation of the QES by its $v$ coordinates affects $\hat I$ but fixes $\tilde I$. In this case, the variation $\delta v_a$ effectively changes a set of modes in $R\cap\tilde I$ for those which include all 3 subsets of modes. The change in the latter does not change the entropy because the state is pure for this set of modes. Hence, the entropy gradient is
\EQ{
\frac{S_\text{QFT}(R\cup I)}{\partial v_a}=\frac{\partial\tilde S(I)}{\partial v_a}-\dot S_\text{rad}^{\mathbb R}=\dot S_\text{rad}-\dot S_\text{rad}^{\mathbb R}\ .
\label{pou}
}

Putting together these results for the variation of the generalized entropy, gives us a pair of equations for $\lambda_a$ and $\mu_a$:
\EQ{
\frac{4\pi{\cal N}T\lambda_a\mu_a}s+\big(\dot S_\text{rad}^{\mathbb R}-\dot S_\text{rad}^{\mathbb T}-\dot S_\text{rad}\big)\Theta(\lambda_a-1)&=0\ ,\\[5pt]
-\frac{4\pi{\cal N}T\lambda_a\mu_a}s+\dot S_\text{BH}+\dot S_\text{rad}-\dot S_\text{rad}^{\mathbb R}&=0\ .
}
The second equation determines the product $\lambda_a\mu_a$. Then the first equations determines $\lambda_a$ as the solution of
\EQ{
\Theta(\lambda_a-1)=\frac{\dot S_\text{rad}-\dot S_\text{rad}^{\mathbb R}+\dot S_\text{BH}}{\dot S_\text{rad}-\dot S_\text{rad}^{\mathbb R}+\dot S_\text{rad}^{\mathbb T}}\ .
\label{cuy}
}
This will have a solution as long as the right-hand side is between 0 and 1. Since $\dot S_\text{BH}<0$ and $\dot S_\text{rad}^{\mathbb T}>0$ and the denominator in \eqref{cuy} is always positive, this will be be true as long as the numerator is positive, that is
\EQ{
\dot S_\text{rad}-\dot S_\text{rad}^{\mathbb R}+\dot S_\text{BH}>0\ .
\label{puy}
}
Note that  $s$ does not appear in the equation for $\lambda_a$ hence the solution has $\lambda_a$ of order $s^0$ and $\mu_a$ of order $s$ as we hypothesized in \eqref{cyr}. 

One can check that \eqref{puy} is always satisfied because the each term is expressed as an integral over $\omega$ and the combined integrand is manifestly positive for any $\Gamma(\omega)$. Equality is approached in the limit  where the transmission is very small, i.e.~$\dot S_\text{BH}\to0$, $\dot S_\text{rad}^{\mathbb  T}\to0$ and $\dot S^{\mathbb  R}_\text{rad}\to \dot S_\text{rad}$.

This completes our proof that extrema of the generalized entropy exist with the pattern we hypothesized.

\section{Nested Page's theorem and multiple saddles}\label{s5}

In sections \ref{s5} and \ref{s6} we apply the formalism we have established, summed up in the formula \eqref{ger2}, to 
phenomena involving the correlations of the Hawking radiation for which the dependence on the grey-body coefficient $\xi$ plays an important role. 

In this section, we investigate a characteristic feature of the islands-in-the-stream formalism in that there is generally a competition from many possible saddles whereas the usual expectation of Page's theorem \cite{Page:1993df} is that there would be a competition between two saddles. 

We will show that the existence of multiple saddles is intrinsically linked to the irreversible nature of evaporation, in the sense that in the reversible limit $\xi\to1$ two saddles can actually dominate and so a na\"\i ve application of Page's theorem is valid. However, when $\xi>1$, we show that there is a nested generalization of Page's theorem that describes the multiple saddles that can dominate.

\subsection{Page curve}

The island prescription is known to reproduce the Page curve for the emitted radiation, consistent with unitarity \cite{Penington:2019npb,Almheiri:2019psf}. This can be easily checked using \eqref{ger2}. Consider an interval of radiation $R=[0,u]$ and for simplicity the case $S_*=0$, so there is no extremal entropy. In this case, there are two possible entropy saddles which can be visualized in terms of the subsets $A$ and the image of the  island-in-the-stream $\tilde I(=\hat I)$ at $\mathscr I^+$ as
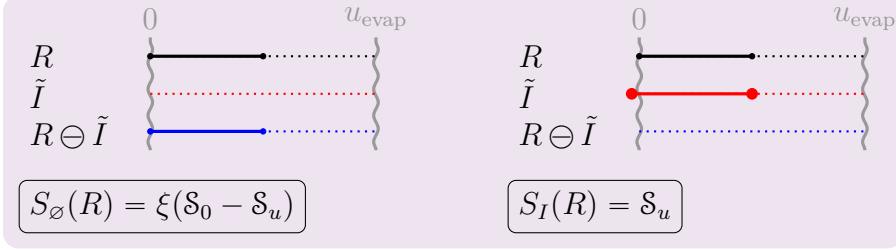
\begin{center}
	\begin{tikzpicture} [scale=0.5]
	\filldraw[fill = Plum!10!white, draw = Plum!10!white, rounded corners = 0.2cm] (-3.9,1.6) rectangle (20,-5.1);
	\draw[decorate,very thick,black!40,decoration={snake,amplitude=0.03cm}] (6,-2.5) -- (6,0.5);
	\draw[decorate,very thick,black!40,decoration={snake,amplitude=0.03cm}] (0,-2.5) -- (-0,0.5);
	\draw[dotted,thick] (0,0) -- (6,0);
	\draw[dotted,thick,red] (0,-1) -- (6,-1);
	\draw[dotted,thick,blue] (0,-2) -- (6,-2);
	\draw[very thick] (0,0) -- (3,0);
	\filldraw[black] (0,0) circle (2pt);
	\filldraw[black] (3,0) circle (2pt);
	\draw[very thick,blue] (0,-2) -- (3,-2);
	\filldraw[blue] (0,-2) circle (2pt);
	\filldraw[blue] (3,-2) circle (2pt);
	\node[right] at (-3.5,0) {$R$};
	\node[right] at (-3.5,-1) {$\tilde I$};
	\node[right] at (-3.5,-2) {$R\ominus\tilde I$};
	\node[right,draw=black,rounded corners=3pt] at (-3.5,-4) {$S_\varnothing(R)=\xi(\ZZ_0-\ZZ_u)$};
	\node[black!40] at (0,1) {$0$};
	\node[black!40] at (6,1) {$u_\text{evap}$};
	\begin{scope}[xshift=13cm]
	%
	%\filldraw[fill = Plum!10!white, draw = Plum!10!white, rounded corners = 0.2cm] (-3.9,1.6) rectangle (14.2,-5.1);
	%
	\draw[decorate,very thick,black!40,decoration={snake,amplitude=0.03cm}] (6,-2.5) -- (6,0.5);
	\draw[decorate,very thick,black!40,decoration={snake,amplitude=0.03cm}] (0,-2.5) -- (-0,0.5);
	\draw[dotted,thick] (0,0) -- (6,0);
	\draw[dotted,thick,red] (0,-1) -- (6,-1);
	\draw[dotted,thick,blue] (0,-2) -- (6,-2);
	\draw[very thick] (0,0) -- (3,0);
	\filldraw[black] (0,0) circle (2pt);
	\filldraw[black] (3,0) circle (2pt);
	\draw[very thick,red] (-0.2,-1) -- (3,-1);
	\filldraw[red] (-0.2,-1) circle (4pt);
	\filldraw[red] (3,-1) circle (4pt);
	\node[right] at (-3.5,0) {$R$};
	\node[right] at (-3.5,-1) {$\tilde I$};
	\node[right] at (-3.5,-2) {$R\ominus\tilde I$};
	\node[right,draw=black,rounded corners=3pt] at (-3.5,-4) {$S_{I}(R)=\ZZ_u$};
	\node[black!40] at (0,1) {$0$};
	\node[black!40] at (6,1) {$u_\text{evap}$};
	\end{scope}
	\end{tikzpicture}
\end{center}
On the left we have the Hawking, or no-island, saddle, for which the entropy is simply
\EQ{
	S_\varnothing(R)=\xi(\ZZ_0-\ZZ_u)\equiv S_\text{rad}^\mathbb{T}(0,u)\ ,
}
where $\ZZ_u\equiv S_\text{BH}(u)$ is the black hole entropy while $\xi$ is the grey-body coefficient, the ratio of the entropy fluxes of the black hole and the radiation \eqref{rry}.

The other possible entropy saddle has an island whose image on $\mathscr I^+$ covers the whole of $R$, i.e.~$\tilde I=[0^-,u]$. In this case, the island purifies all the transmitted radiation $S_\text{rad}^{\mathbb T} (R\ominus \tilde{I})=0$ and so the only contribution comes from the QES at $u$:
\EQ{
	S_I(R)=\ZZ_u\equiv S_\text{BH}(u)\ .
}
Note that the contribution from the QES at $u=0^-$ vanishes.

The entropy of $R$ is then the minimum among these two contributions
\EQ{
	S(R)=\min\big(S_\text{rad}^{\mathbb T}(0,u),S_\text{BH}(u)\big)\ ,
	\label{eq:Page_curve}
}
which is the Page curve. The transition between the two saddles occurs at the Page time
\EQ{
	\ZZ_\text{Page}=\frac\xi{\xi+1}\ZZ_0\ .
	\label{nur}
}
Notice that in the reversible case $\xi\to1$  the Page transition occurs when the black hole has lost exactly half its entropy. 

The result \eqref{eq:Page_curve} is just a statement of Page's theorem, see equation \eqref{eq:page_theorem}, if we identify the effective Hilbert space dimension of the radiation and the black hole as
\EQ{
S^{\mathbb T}_\text{rad}(0,u)\thicksim\log d_R\ ,\qquad S_\text{BH}(u)\thicksim\log d_B\ .
}
One route to Page's theorem \cite{Page:1993df} is to take a bipartite quantum system ${\cal H}_R\otimes{\cal H}_B$ and compute the R\'enyi entropy $S^{(n)}(R)$ of subsystem $R$ of some pure state of the total system $\rho_0=\ket{\psi_0}\bra{\psi_0}$ and then average over the pure state. Conceptually the idea is analogous to statistical mechanics where an ensemble average captures the behaviour of a single system in a random, or typical, pure state. 

In the present context, one actually averages the exponential of the R\'enyi entropies
\EQ{
e^{(1-n)S^{(n)}(R)}=\int \tr_R\big[\rho_R(U)^n\big]\,dU\ ,\qquad\text{where}\quad \rho_R(U)=\tr_B\big(U\rho_0U^\dagger\big)\ ,
}
where $U$ is an element of the group $SU(N)$, $N=d_Rd_B$, and the measure is the standard group invariant measure. These kinds of averages can be computed by introducing replicas for the Hilbert space ${\cal H}\to{\cal H}^{\otimes n}$:
\EQ{
\int\tr_R\big[\rho_R(U)^n\big]\,dU&=\int\tr^{(n)}\big[(U\rho_0 U^\dagger)^{\otimes n}\eta_R\otimes e_B\big]\,dU\\[5pt] &=
\int\tr^{(n)}\big[\rho_0^{\otimes n}(U^\dagger)^{\otimes n}\eta_R\otimes e_BU^{\otimes n}\big]\,dU
\ ,
\label{klk}
}
where $\eta_R$ is the cyclic permutation on the $n$ copies of the subsystem $R$, $e_B$ is the identity permutation on $B$ and the trace is taken over the $n$ replicas of the total space ${\cal H}_R\otimes{\cal H}_B$. In the second equality we have used the cyclicity of the trace because we find it more convenient to average the permutation rather than the state. When $N\gg1$, then the average is dominated by a sum of terms that are associated to elements of the symmetric group $S_n$ acting on the replicas:\footnote{The average is invariant under the action of $\text{SU}(N)$ acting diagonally on the $n$ replicas and this means that, as a consequence of Schur-Weyl duality, it must be a linear combination of elements of its commutant, the symmetric group $S_n$ acting on the replicas. The precise formula on the right-hand side is $\sum_{\tau,\sigma\in S_n}\text{Wg}(\sigma^{-1}\tau,N)\tr(\tau^{-1} f)\sigma$, involving the Weingarten function. When $N$ is large the term with $\sigma=\tau$ dominates and $\text{Wg}(1,N)=1/N^n+\cdots$. }
\EQ{
\llbracket f\rrbracket\equiv\int (U^\dagger)^{\otimes n}\,f\, U^{\otimes n}\,dU\approx\frac1{N^n}\sum_{\tau\in S_n}\tr^{(n)}(\tau^{-1} f)\tau\ .
\label{vuu}
}
Hence, we have
\EQ{
\llbracket\eta_R\otimes e_B\rrbracket\approx\frac1{N^n}\sum_{\tau\in S_n}\tr^{(n)}(\tau^{-1}\eta_R\otimes e_B)\tau\ .
}
Then we have $\tr^{(n)}(\tau^{-1}\eta_R\otimes e_B)=d_R^{k(\eta^{-1}\tau)}d_B^{k(\tau)}$, where $k(\tau)$ is the number of cycles in the element $\tau$, so $k(e)=n$ and $k(\eta)=1$. Finally, since $\rho_0$ is pure, $\tr^{(n)}(\rho_0^{\otimes n}\tau)=1$ for any $\tau$, and so
\EQ{
e^{(1-n)S^{(n)}(R)}\approx\frac1{N^n}\sum_{\tau\in S_n}d_R^{k(\eta^{-1}\tau)}d_B^{k(\tau)}\ .
\label{zoj}
}
The sum is dominated by two terms corresponding to the identity $e$ and the cyclic permutation $\eta$, hence
\EQ{
e^{(1-n)S^{(n)}(R)}=d_R^{1-n}+d_B^{1-n}+\cdots
}
and so $S(R)=\lim_{n\to1}S^{(n)}(R)\approx\min(\log d_R,\log d_B)$. The other terms in \eqref{zoj} in the sum have the effect of smoothing over the crossover when $d_R\sim d_B$.

\subsection{Three competing saddles}\label{s5.2}

Now consider the case when the interval $R$ no longer begins at $u=0$, rather $R=[u_1,u_2]$ with $u_1>0$. According to the islands-in-the-stream formalism, there are now three saddles that can contribute \cite{Hollowood:2021nlo}:
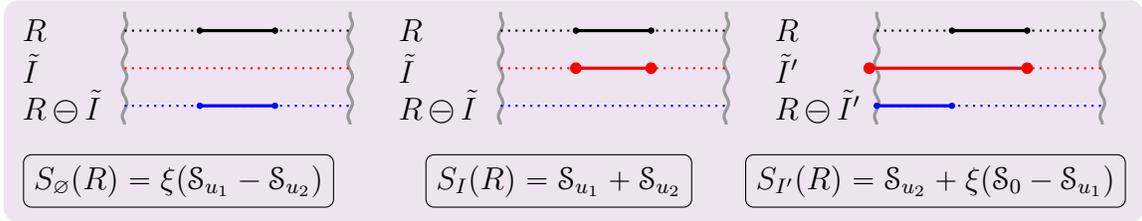
\begin{center}
	\begin{tikzpicture} [scale=0.5]
	\filldraw[fill = Plum!10!white, draw = Plum!10!white, rounded corners = 0.2cm] (-3.2,0.8) rectangle (27.1,-5.1);
	\draw[decorate,very thick,black!40,decoration={snake,amplitude=0.03cm}] (6,-2.5) -- (6,0.5);
	\draw[decorate,very thick,black!40,decoration={snake,amplitude=0.03cm}] (0,-2.5) -- (0,0.5);
	\draw[dotted,thick] (0,0) -- (6,0);
	\draw[dotted,thick,red] (0,-1) -- (6,-1);
	\draw[dotted,thick,blue] (0,-2) -- (6,-2);
	%\draw[thick,dashed] (-4,2) -- (1,7);
	\draw[very thick] (2,0) -- (4,0);
	\filldraw[black] (2,0) circle (2pt);
	\filldraw[black] (4,0) circle (2pt);
	\draw[very thick,blue] (2,-2) -- (4,-2);
	\filldraw[blue] (2,-2) circle (2pt);
	\filldraw[blue] (4,-2) circle (2pt);
	\node[right,draw=black,rounded corners=3pt] at (-2.7,-4) {$S_\varnothing(R)=\xi(\ZZ_{u_1}-\ZZ_{u_2})$};
	\node[right] at (-3,0) {$R$};
	\node[right] at (-3,-1) {$\tilde I$};
	\node[right] at (-3,-2) {$R\ominus\tilde I$};
	\begin{scope}[xshift=10cm]
	\draw[decorate,very thick,black!40,decoration={snake,amplitude=0.03cm}] (6,-2.5) -- (6,0.5);
	\draw[decorate,very thick,black!40,decoration={snake,amplitude=0.03cm}] (0,-2.5) -- (0,0.5);
	\draw[dotted,thick] (0,0) -- (6,0);
	\draw[dotted,thick,red] (0,-1) -- (6,-1);
	\draw[dotted,thick,blue] (0,-2) -- (6,-2);
	%\draw[thick,dashed] (-4,2) -- (1,7);
	\draw[very thick] (2,0) -- (4,0);
	\filldraw[black] (2,0) circle (2pt);
	\filldraw[black] (4,0) circle (2pt);
	\draw[very thick,red] (2,-1) -- (4,-1);
	\filldraw[red] (2,-1) circle (4pt);
	\filldraw[red] (4,-1) circle (4pt);
	\node[right,draw=black,rounded corners=3pt] at (-2,-4) {$S_I(R)=\ZZ_{u_1}+\ZZ_{u_2}$};
	\node[right] at (-3,0) {$R$};
	\node[right] at (-3,-1) {$\tilde I$};
	\node[right] at (-3,-2) {$R\ominus\tilde I$};
	\end{scope}
	\begin{scope}[xshift=20cm]
\draw[decorate,very thick,black!40,decoration={snake,amplitude=0.03cm}] (6,-2.5) -- (6,0.5);
\draw[decorate,very thick,black!40,decoration={snake,amplitude=0.03cm}] (0,-2.5) -- (0,0.5);
\draw[dotted,thick] (0,0) -- (6,0);
\draw[dotted,thick,red] (0,-1) -- (6,-1);
\draw[dotted,thick,blue] (0,-2) -- (6,-2);
\draw[very thick] (2,0) -- (4,0);
\filldraw[black] (2,0) circle (2pt);
\filldraw[black] (4,0) circle (2pt);
\draw[very thick,red] (-0.2,-1) -- (4,-1);
\filldraw[red] (-0.2,-1) circle (4pt);
\filldraw[red] (4,-1) circle (4pt);
\draw[very thick,blue] (0,-2) -- (2,-2);
\filldraw[blue] (0,-2) circle (2pt);
\filldraw[blue] (2,-2) circle (2pt);
\node[right,draw=black,rounded corners=3pt] at (-3.5,-4) {$S_{I'}(R)=\ZZ_{u_2}+\xi(\ZZ_0-\ZZ_{u_1})$};
\node[right] at (-3,0) {$R$};
\node[right] at (-3,-1) {$\tilde I'$};
\node[right] at (-3,-2) {$R\ominus\tilde I'$};
	\end{scope}
	\end{tikzpicture}
\end{center}
\noindent  Hence,
\EQ{
S(R)= \min\big(S_\text{rad}^{\mathbb T}(u_1,u_2),S_\text{BH}(u_1) + S_\text{BH}(u_2), S_\text{rad}^{\mathbb T}(0,u_1)+S_\text{BH}(u_2)\big)\ .
\label{kws}
}

It is immediately apparent that the saddle with island $I$ can only dominate in the irreversible situation $\xi>1$. In the reversible case, there is a competition of two saddles $\emptyset$ and $I'$ only and the relation to Page's theorem is then immediate. In fact, one can show that this reduction to only two saddles occurs for any 
general subset $R=[u_1,u_2]\cup\cdots\cup[u_{2p-1},u_{2p}]$ of the radiation. Using equation \eqref{eq:gen3}, the two saddles that can compete are the Hawking saddle,
\EQ{
S_\varnothing(R)=S_\text{rad}^{\mathbb T}(R)=\sum_{j=1}^{2p}(-1)^{j+1}\ZZ_j\ ,
}
and the island saddle with island-in-the-stream $\tilde I=[0^-,u_{2p}]$, with entropy
\EQ{
S_I(R)=S_\text{rad}^{\mathbb T}(\overline R)+S_\text{BH}(u_{2p})=\sum_{j=0}^{2p}(-1)^j\ZZ_j\ ,
}
where $\overline R=[0,u_1]\cup\cdots\cup[u_{2p-2},u_{2p-1}]\subset\mathscr I^+$ is complementary to $R$ up to time $u_{2p}$. So the entropy is then a competition between these two saddles
\EQ{
S(R)=\min\big(S_\text{rad}^{\mathbb T}(R),S_\text{rad}^{\mathbb T}(\overline R)+S_\text{BH}(u_{2p})\big)\ ,
}
in accordance with Page's theorem. 

\begin{figure}
\begin{center}
\begin{tikzpicture} [scale=0.8]
\draw[fill = Plum!10!white,line width=1mm,rounded corners = 0.2cm] (-1,-1) rectangle (1,1);
\draw[decorate,line width=0.6mm,decoration={snake,amplitude=0.03cm},-triangle 60] (1,0.5) -- (3.5,1.5) node[sloped,midway,above] {$\overline R$};
\draw[-triangle 60,line width=0.6mm] (-3.5,0) -- (-1,0) node[sloped,midway,above] {$B_0$};
%\draw[-triangle 60,line width=0.6mm] (-2,-1.5) -- (-2,-0.5)  -- (-1,-0.5);
%
\draw[-triangle 60,line width=0.6mm] (1,-0.5)  -- (3.5,-0.5) node[sloped,midway,above] {$B_1$};
%
%\draw [thick,decorate,decoration={brace,amplitude=10pt}] (-3.7,-1.7) -- (-3.7,0.2) node [black,midway,xshift=-0.8cm] {$\rho_0$};
%
%\node at (-2,-2.1) {$Q_1$};
%
\node at (0,0) {\Large $U_1$};
\begin{scope}[xshift=0.5cm]
\draw[fill = Plum!10!white,line width=1mm,rounded corners = 0.2cm] (3,-2) rectangle (5,0);
\draw[decorate,line width=0.6mm,decoration={snake,amplitude=0.03cm},-triangle 60] (5,-0.5) -- (7.5,0.5) node[sloped,midway,above] {$R$};
\draw[-triangle 60,line width=0.6mm] (0.5,-2) -- (3,-1.5) node[sloped,midway,below] {$Q_1$};
\draw[-triangle 60,line width=0.6mm] (5,-1.5) -- (7.5,-1.5) node[sloped,midway,above] {$B_2$};
\node at (4,-1) {\Large $U_2$};
%
%\node at (-2.8,-1) {$Q_2$};
%
\end{scope}
\end{tikzpicture}
\caption{\footnotesize A model for the black hole evaporation for the evaluation of the entropy of the subset of radiation $S(R)$. The $B_i$ are the black hole at times $0$, $u_1$ and $u_2$. In order to model irreversible evaporation an auxiliary factor $Q_1$ must be added in as indicated.}
\label{fig4}
\end{center}
\end{figure}
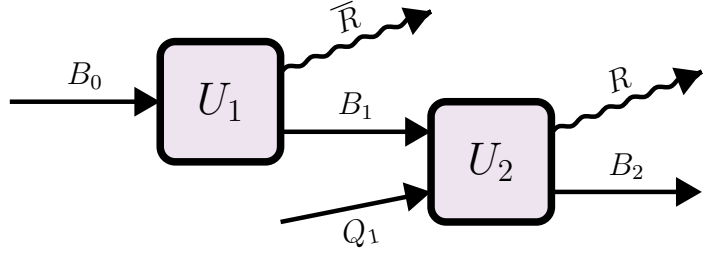

Now we return to the case of irreversible evaporation $\xi>1$ and consider how we can square the appearance of three competing saddles with Page's theorem. The reason why Page's theorem na\"\i vely does not apply is subtle but boils down to the fact that in the irreversible case, the overall Hilbert space is effectively getting bigger during the evaporation. To take account of this we need to add in some auxiliary Hilbert space factors as the evaporation proceeds \cite{Penington:2019npb,Hayden:2018khn}. The upshot is that there is a nested construction that leads to a generalization of Page's theorem. The set up is shown in figure \ref{fig4}. The auxiliary Hilbert space factor needed to account for the irreversibility  is labelled $Q_1$ and the initial pure state is an unentangled state $\ket{\psi_0}=\ket{\alpha}_{B_0}\otimes\ket{\beta}_{Q_1}$.

The averaging procedure now involve two separate unitary averages that act on the factors as shown in the figure. The R\'enyi entropies of the subset $R$ are given by 
\EQ{
e^{(1-n)S^{(n)}(R)}&=\int \tr_R\big(\tr_{\overline RB_2}U_2U_1\rho_0U_1^\dagger U_2^\dagger\big)^n\,dU_1\,dU_2\\
&=\tr^{(n)}\big(\rho_0^{\otimes n}\llbracket e_{\overline R}\otimes \llbracket \eta_R\otimes e_{B_2} \rrbracket_2\rrbracket_1\big)\ ,
}
where in the last line we introduced replicas and used the cyclicity of the trace. We can then compute the averages in turn, using \eqref{vuu}, picking out only the terms that can dominate when the Hilbert space factors all have a large dimension,
\EQ{
\llbracket\eta_R\otimes e_{B_2}\rrbracket_2=d_R^{1-n}e_{B_1}\otimes e_{Q_1}+d_{B_2}^{1-n}\eta_{B_1}\otimes \eta_{Q_1}+\cdots
}
and then 
\EQ{
\llbracket\eta_{B_1}\otimes e_{\overline R}\rrbracket_1=d_{B_1}^{1-n}e_{B_0}+d_{\overline R}^{1-n}\eta_{B_0}+\cdots\ ,\qquad \llbracket e_{B_1}\otimes e_{\overline R}\rrbracket_1=1\ .
} 
Assembling all the pieces, and using the fact that for the final trace $\tr^{(n)}(\rho_0^{\otimes n}\tau_{B_0}\otimes\sigma_{Q_1})=1$, for any $\tau$ and $\sigma$, gives
\EQ{
e^{(1-n)S^{(n)}(R)}=d_R^{1-n}+(d_{B_1}d_{B_2})^{1-n}+(d_{\overline R}d_{B_2})^{1-n}+\cdots\ .
}
Away from the crossover regions, this gives the von Neumann entropy 
\EQ{
S(R)=\min\big(\log d_R,\log(d_{B_1}d_{B_2}),\log(d_{\overline R}d_{B_2})\big)\ .
}
Given the identification $S_\text{rad}(u_1,u_2)\thicksim \log d_R$, $S_\text{rad}(0,u_1)\thicksim \log d_{\overline R}$ and $S_\text{BH}(u_j)\thicksim\log d_{B_j}$, the result here is precisely competition between the three saddles in \eqref{kws}. Note that the r\^ole of the auxiliary factor $Q_1$ is simply to allow $d_Rd_{B_2}>d_{B_1}$, that is 
\EQ{
S^{\mathbb T}_\text{rad}(u_1,u_2)> S_\text{BH}(u_1)-S_\text{BH}(u_2)\ .
}
More precisely, we have
\begin{equation}
\xi = 1 + \frac{\log d_{Q_1}}{\log d_{B_1}/d_{B_2}}
\end{equation}
which means $\xi > 1$ and the reversible limit is obtained when $d_{Q_1}=1$ .

What is interesting is that there is a concrete relation between the elements $\tau_j$, $j=1,2$, of the symmetric groups $S_n$, which act on replicas of subsystems $B_0$ and $B_1\cup Q_1$, respectively,  associated to each of the averages over the unitary group, as in \eqref{vuu}, and the islands-in-the-stream as shown in the table:
\begin{center}
\begin{tabular}{lcll}
\toprule
\multicolumn{2}{c} {nested model} &\multicolumn{2}{c}{islands} \\[3pt]
\multicolumn{1}{c}{$S_{\{\tau_j\}}(R)$} & \multicolumn{1}{c}{$(\tau_1,\tau_2)$} & \multicolumn{1}{c}{$S_I(R)$} & \multicolumn{1}{c}{$\tilde I$} \\ \midrule 
\rowcolor{Cyan!15} $\log d_{R}$ & $(e,e)$ & $S^{\mathbb T}_\text{rad}(u_1,u_2)$  & $\emptyset$  \\
\rowcolor{Melon!15} $\log(d_{B_1}d_{B_2})$ & $(e,\eta)$ & $S_\text{BH}(u_1)+S_\text{BH}(u_2)$  & $R$  \\ 
\rowcolor{Cyan!15} $\log(d_{\overline R}d_{B_2})$  & $(\eta,\eta)$ & $S_\text{BH}(u_2)+S^{\mathbb T}_\text{rad}(0,u_1)$ & $\overline R\cup R$  \\  \bottomrule
\end{tabular}
\end{center}
Notice that not all combinations of $(\tau_1,\tau_2)$, with $\tau_j\in\{e,\eta\}$, actually occur, in particular, the combination $(\eta,e)$ cannot occur. We will sum up this in effective selection rules in the next section. It is also clear from the table that there is a relation between the island-in-the-stream and the string of elements of $e,\eta\in S_n$, again this will emerge in generality in the next section.

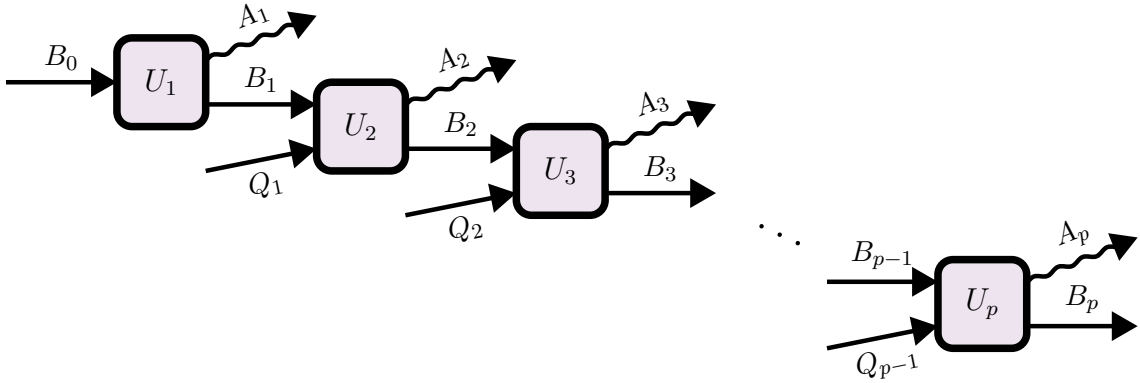
\begin{figure}
\begin{center}
\begin{tikzpicture} [scale=0.59]
\draw[fill = Plum!10!white,line width=1mm,rounded corners = 0.2cm] (-1,-1) rectangle (1,1);
\draw[decorate,line width=0.6mm,decoration={snake,amplitude=0.03cm},-triangle 60] (1,0.5) -- (3.5,1.5) node[sloped,midway,above] {\small$A_1$};
\draw[-triangle 60,line width=0.6mm] (-3.5,0) -- (-1,0) node[sloped,midway,above] {\small$B_0$};
\draw[-triangle 60,line width=0.6mm] (1,-0.5)  -- (3.5,-0.5) node[sloped,midway,above] {\small$B_1$};
\node at (0,0) {$U_1$};
\begin{scope}[xshift=0.5cm]
\draw[fill = Plum!10!white,line width=1mm,rounded corners = 0.2cm] (3,-2) rectangle (5,0);
\draw[decorate,line width=0.6mm,decoration={snake,amplitude=0.03cm},-triangle 60] (5,-0.5) -- (7.5,0.5) node[sloped,midway,above] {\small$A_2$};
\draw[-triangle 60,line width=0.6mm] (0.5,-2) -- (3,-1.5) node[sloped,midway,below] {\small$Q_1$};
\draw[-triangle 60,line width=0.6mm] (5,-1.5) -- (7.5,-1.5) node[sloped,midway,above] {\small$B_2$};
\node at (4,-1) {$U_2$};
\end{scope}
\begin{scope}[xshift=5cm,yshift=-1cm]
\draw[fill = Plum!10!white,line width=1mm,rounded corners = 0.2cm] (3,-2) rectangle (5,0);
\draw[decorate,line width=0.6mm,decoration={snake,amplitude=0.03cm},-triangle 60] (5,-0.5) -- (7.5,0.5) node[sloped,midway,above] {\small$A_3$};
\draw[-triangle 60,line width=0.6mm] (0.5,-2) -- (3,-1.5) node[sloped,midway,below] {\small$Q_2$};
\draw[-triangle 60,line width=0.6mm] (5,-1.5) -- (7.5,-1.5) node[sloped,midway,above] {\small$B_3$};
\node at (4,-1) {$U_3$};
\end{scope}
\begin{scope}[xshift=14.5cm,yshift=-4cm]
\draw[fill = Plum!10!white,line width=1mm,rounded corners = 0.2cm] (3,-2) rectangle (5,0);
\draw[decorate,line width=0.6mm,decoration={snake,amplitude=0.03cm},-triangle 60] (5,-0.5) -- (7.5,0.5) node[sloped,midway,above] {\small$A_p$};
\draw[-triangle 60,line width=0.6mm] (0.5,-0.5) -- (3,-0.5) node[sloped,midway,above] {\small$B_{p-1}$};
\draw[-triangle 60,line width=0.6mm] (0.5,-2) -- (3,-1.5) node[sloped,midway,below] {\small$Q_{p-1}$};
\draw[-triangle 60,line width=0.6mm] (5,-1.5) -- (7.5,-1.5) node[sloped,midway,above] {\small$B_p$};
\node at (4,-1) {$U_p$};
\end{scope}
\node[rotate=-25] at (14,-3.5) {\Large$\cdots$};
\end{tikzpicture}
\caption{\footnotesize The general case involves a nested series of unitary averages and Hilbert space factors as shown.}
\label{fig8}
\end{center}
\end{figure}

\subsection{General radiation region $R$}

The analysis can easily be extended to cover any general subset of the radiation $R$. Let us define this in the following way. Define subsets $A_j=[u_{j-1},u_j]$, $j=1,2\ldots,p$, with $u_0=0$ and where $|u_j-u_{j-1}|\gg\Delta t_s$,\footnote{If we want to mimic an evaporation process which happens at discrete time steps, we can choose the $u_j$ to be equidistant.} then $R$ can be taken to be a subset of $\{A_j\}$. The set up is shown schematically in figure \ref{fig8}. The general expression for the R\'enyi entropy is given by
\begin{equation}
e^{(1-n)S^{(n)}(R)} = \tr^{(n)}\big(\rho_0^{\otimes n}\llbracket \tau_{A_1} \otimes \dots \llbracket \tau_{A_{p-1}}\otimes \llbracket \tau_{A_p} \otimes e_{B_p} \rrbracket_p\rrbracket_{p-1} \dots \rrbracket_1\big)
\label{eq:general_R_qubit}
\end{equation} 
where $\tau_{A_j}$ can be $\eta_{A_j}$ or $e_{A_j}$ depending on whether $A_j$ is or is not in $R$ respectively, and we have used \eqref{vuu} and the cyclicity of the trace $p$ times, one for each unitary averages.
This means that we have $p$ elements of the permutation group $\tau_j$, $j=1,2,\ldots,p$, which act on the replicas of ${\cal H}_{B'_{j-1}}\equiv{\cal H}_{B_{j-1}}\otimes{\cal H}_{Q_{j-1}}$. 
As we have seen in the previous section, the dominant contributions in \eqref{eq:general_R_qubit} occur when each element $\tau_j$ is either $e$ or $\eta$. The rules follow from the averages 
\EQ{
\llbracket e_{A_j}\otimes e_{B_j}\rrbracket_j&=e_{B'_{j-1}}\ ,\qquad
\llbracket \eta_{A_j}\otimes \eta_{B_j}\rrbracket_j=\eta_{B'_{j-1}}\ ,\\[8pt]
\llbracket \eta_{A_j}\otimes e_{B_j}\rrbracket_j&=d_{A_j}^{1-n}\,e_{B'_{j-1}}+d_{B_j}^{1-n}\,\eta_{B'_{j-1}}\ ,\\[8pt]
\llbracket e_{A_j}\otimes \eta_{B_j}\rrbracket_j&=d_{B_j}^{1-n}\,e_{B'_{j-1}}+d_{A_j}^{1-n}\,\eta_{B'_{j-1}}\ ,
\label{bbo}
}
where we have not considered subleading terms associated to other elements of $S_n$. Inserting \eqref{bbo} in \eqref{eq:general_R_qubit} one can explicitly recover \eqref{eq:gen3} after the identification $\log d_{A_j} = \xi (\ZZ_{u_j}-\ZZ_{u_{j-1}})$ and $\log d_{B_j} = \ZZ_{u_j}$. Let's explore a bit more this identification.

As we saw in the previous section, not every combination of elements can occur: there are effectively selection rules between an adjacent pair $(\tau_j,\tau_{j+1})$.
In equation \eqref{bbo}, the contributions $\tau_{B'_{j-1}}$, with $\tau\in\{e,\eta\}$, on the right-hand side are identified with $\tau_j$, while the left-hand side involves $\tau_{j+1}$ acting on $B_j$. The first line in \eqref{bbo} manifests the effective selection rules on $(\tau_j,\tau_{j+1})$: $(\eta,e)$ cannot occur if $A_j \not\in R$, i.e. for $e_{A_j}$, while $(e,\eta)$ cannot occur if $A_j \in R$, i.e. for $\eta_{A_j}$. This two rules combined together are equivalent to the claim, in the island in the stream formula \eqref{ger2}, that the QES image on $\mathscr{I}^+$ must correspond to the radiation endpoints.

\begin{figure}
\begin{center}
	\begin{tikzpicture} [scale=0.8]
	\filldraw[fill = Plum!10!white, draw = Plum!10!white, rounded corners = 0.2cm] (-2.8,1.6) rectangle (9.7,-4.7);
	\draw[decorate,very thick,black!40,decoration={snake,amplitude=0.03cm}] (0,-4.2) -- (0,1);
	\draw[dotted,thick] (0,0) -- (9,0);
	\draw[dotted,thick,red] (0,-1) -- (9,-1);
	\draw[dotted,thick,blue] (0,-2) -- (9,-2);
	\draw[very thick] (6,0) -- (9,0);
	\draw[very thick] (2,0) -- (4,0);	
	\draw[very thick,red] (2,-1) -- (9,-1);
\draw[black!40,dashed] (1,1) -- (1,-4.2);
\draw[black!40,dashed] (2,1) -- (2,-4.2);
\draw[black!40,dashed] (3,1) -- (3,-4.2);
\draw[black!40,dashed] (4,1) -- (4,-4.2);
\draw[black!40,dashed] (5,1) -- (5,-4.2);
\draw[black!40,dashed] (6,1) -- (6,-4.2);
\draw[black!40,dashed] (7,1) -- (7,-4.2);
\draw[black!40,dashed] (8,1) -- (8,-4.2);
\draw[black!40,dashed] (9,1) -- (9,-4.2);			
	\filldraw[black] (6,0) circle (2pt);
	\filldraw[black] (9,0) circle (2pt);
	\filldraw[black] (2,0) circle (2pt);
	\filldraw[black] (4,0) circle (2pt);	
	\filldraw[red] (2,-1) circle (4pt);
	\filldraw[red] (9,-1) circle (4pt);	
	\draw[very thick,blue] (4,-2) -- (6,-2);
	\filldraw[blue] (4,-2) circle (2pt);
	\filldraw[blue] (6,-2) circle (2pt);
	\node[right] at (-2.5,0) {$R$};
	\node[right] at (-2.5,-1) {$\tilde I$};
	\node[right] at (-2.5,-2) {$R\ominus\tilde I$};
	\node[right] at (-2.5,-3) {$\tau_j$};	
	\node[right] at (-2.5,-4) {$S_{\{\tau_j\}}(R)$};	
\node at (0.5,1) {$A_1$};
\node at (1.5,1) {$A_2$};
\node at (2.5,1) {$A_3$};
\node at (3.5,1) {$A_4$};
\node at (4.5,1) {$A_5$};
\node at (5.5,1) {$A_6$};
\node at (6.5,1) {$A_7$};
\node at (7.5,1) {$A_8$};
\node at (8.5,1) {$A_9$};
\node at (0.5,-3) {$e$};
\node at (1.5,-3) {$e$};
\node at (0.5,-4) {$0$};
\node at (2.5,-3) {$\eta$};
\node at (1.5,-4) {$b_2$};
\node at (3.5,-3) {$\eta$};
\node at (2.5,-4) {$0$};
\node at (4.5,-3) {$\eta$};
\node at (3.5,-4) {$0$};
\node at (5.5,-3) {$\eta$};
\node at (4.5,-4) {$a_5$};
\node at (6.5,-3) {$\eta$};
\node at (5.5,-4) {$a_6$};
\node at (7.5,-3) {$\eta$};
\node at (6.5,-4) {$0$};
\node at (8.5,-3) {$\eta$};
\node at (7.5,-4) {$0$};
%\node at (9,-3) {$e$};
\node at (8.5,-4) {$b_9$};
%\node at (9,-3) {$\eta$};
	%
	%\node[black!40] at (0,1) {$0$};
	%
	\end{tikzpicture}
\caption{\footnotesize An example of a contribution to the entropy associated to $\{\tau_j\}$. Using the shorthand $a_j=\log d_{A_j}\approx S_\text{rad}^{\mathbb T}(u_{j-1},u_j)$ and $b_j=\log d_{B_j}\approx S_\text{BH}(u_j)$, the entropy is $S_{\{\tau_j\}}(R)=b_2+a_5+a_6+b_9=S_\text{BH}(u_2)+S^{\mathbb T}_\text{rad}(u_4,u_6)+S_\text{BH}(u_9)$ and the figure shows the relation to the islands-in-the-stream formalism.}
\label{fig6}
\end{center}
\end{figure}
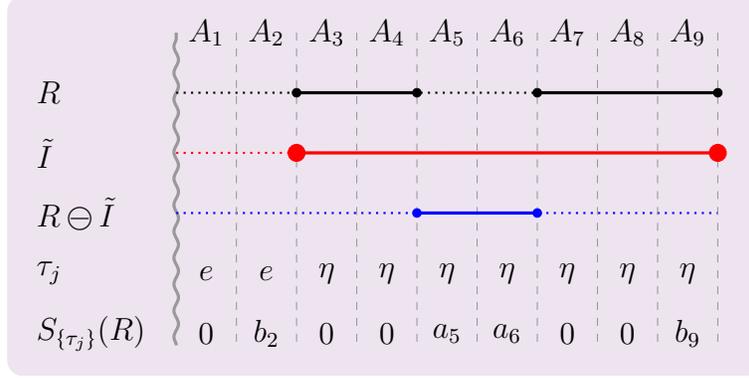

In order to make this last point more explicit, we can consider an example of a particular contribution that corresponds to a saddle is shown in figure \ref{fig6}. It is apparent from this example that, as mentioned above, there is a simple relation between the string of elements of the symmetric group and the island-in-the-stream of the saddle, namely,
\EQ{
S_I(R)\equiv S_{\{\tau_j\}}(R)\qquad\text{where}\qquad\tilde I=\bigcup_{\tau_j=\eta}A_j\ ,
}
What is noteworthy is that the effective selection rules implicit in the averages \eqref{bbo} mesh precisely with the rules of the islands-in-the-stream formalism.

To summarize, we have shown that the island-in-the-stream formalism is captured by a coarse-grained model for the evaporation based on finite dimensional Hilbert space and with a temporal series of unitary ensemble averages. 

\subsection{Decoupling}\label{s5.4}

A simple application of what we have seen above is to study when two intervals of radiation, say $A=A_1\cup A_2$, are not correlated. This happens when the mutual information $I(A_1,A_2)=S(A_1)+S(A_2)-S(A)=0$, or, equivalently, the state on $A$ factorizes $\rho_A=\rho_{A_1}\otimes\rho_{A_2}$. 

In order to investigate this, let us consider once again the set up as in section \ref{s5.2} where $A=\overline R\cup B_2$ is the complementary region to $R$. The relevant contributions to the entropy $S(A)$ are:
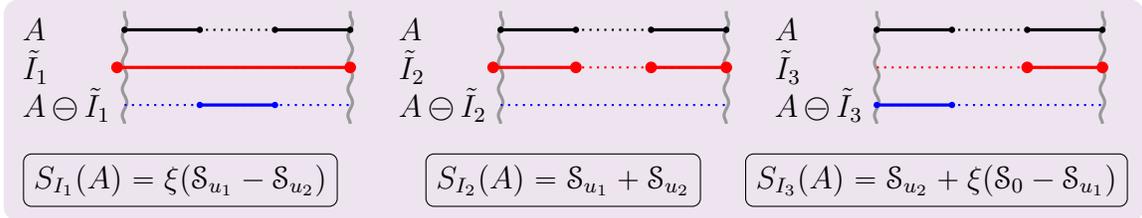
\begin{center}
	\begin{tikzpicture} [scale=0.5]
	\filldraw[fill = Plum!10!white, draw = Plum!10!white, rounded corners = 0.2cm] (-3.2,0.8) rectangle (27.1,-5.1);
	\draw[decorate,very thick,black!40,decoration={snake,amplitude=0.03cm}] (6,-2.5) -- (6,0.5);
	\draw[decorate,very thick,black!40,decoration={snake,amplitude=0.03cm}] (0,-2.5) -- (0,0.5);
	\draw[dotted,thick] (0,0) -- (6,0);
	\draw[dotted,thick,red] (0,-1) -- (6,-1);
	\draw[dotted,thick,blue] (0,-2) -- (6,-2);
	%\draw[thick,dashed] (-4,2) -- (1,7);
	\draw[very thick] (2,0) -- (0,0);
	\draw[very thick] (6,0) -- (4,0);
	\filldraw[black] (2,0) circle (2pt);
	\filldraw[black] (4,0) circle (2pt);
	\filldraw[black] (0,0) circle (2pt);
	\filldraw[black] (6,0) circle (2pt);
	\draw[very thick,red] (-0.2,-1) -- (6,-1);
	\filldraw[red] (-0.2,-1) circle (4pt);
	\filldraw[red] (6,-1) circle (4pt);
	\draw[very thick,blue] (2,-2) -- (4,-2);
	\filldraw[blue] (2,-2) circle (2pt);
	\filldraw[blue] (4,-2) circle (2pt);
	\node[right] at (-3,0) {$A$};
	\node[right] at (-3,-1) {$\tilde I_1$};
	\node[right] at (-3,-2) {$A\ominus\tilde I_1$};
	\node[right,draw=black,rounded corners=3pt] at (-2.7,-4) {$S_{I_1}(A)=\xi(\ZZ_{u_1}-\ZZ_{u_2})$};
	\begin{scope}[xshift=10cm]
	\draw[decorate,very thick,black!40,decoration={snake,amplitude=0.03cm}] (6,-2.5) -- (6,0.5);
	\draw[decorate,very thick,black!40,decoration={snake,amplitude=0.03cm}] (0,-2.5) -- (0,0.5);
	\draw[dotted,thick] (0,0) -- (6,0);
	\draw[dotted,thick,red] (0,-1) -- (6,-1);
	\draw[dotted,thick,blue] (0,-2) -- (6,-2);
	%\draw[thick,dashed] (-4,2) -- (1,7);
	\draw[very thick] (2,0) -- (0,0);
	\draw[very thick] (6,0) -- (4,0);
	\filldraw[black] (2,0) circle (2pt);
	\filldraw[black] (4,0) circle (2pt);
	\filldraw[black] (0,0) circle (2pt);
	\filldraw[black] (6,0) circle (2pt);
	\draw[very thick,red] (2,-1) -- (-0.2,-1);
	\filldraw[red] (2,-1) circle (4pt);
	\filldraw[red] (-0.2,-1) circle (4pt);
	\draw[very thick,red] (6,-1) -- (4,-1);
	\filldraw[red] (6,-1) circle (4pt);
	\filldraw[red] (4,-1) circle (4pt);
	\node[right,draw=black,rounded corners=3pt] at (-2,-4) {$S_{I_2}(A)=\ZZ_{u_1}+\ZZ_{u_2}$};
	\node[right] at (-3,0) {$A$};
	\node[right] at (-3,-1) {$\tilde I_2$};
	\node[right] at (-3,-2) {$A\ominus\tilde I_2$};
	\end{scope}
	\begin{scope}[xshift=20cm]
	\draw[decorate,very thick,black!40,decoration={snake,amplitude=0.03cm}] (6,-2.5) -- (6,0.5);
	\draw[decorate,very thick,black!40,decoration={snake,amplitude=0.03cm}] (0,-2.5) -- (0,0.5);
	\draw[dotted,thick] (0,0) -- (6,0);
	\draw[dotted,thick,red] (0,-1) -- (6,-1);
	\draw[dotted,thick,blue] (0,-2) -- (6,-2);
	\draw[very thick] (2,0) -- (0,0);
	\draw[very thick] (6,0) -- (4,0);
	\filldraw[black] (2,0) circle (2pt);
	\filldraw[black] (4,0) circle (2pt);
	\filldraw[black] (0,0) circle (2pt);
	\filldraw[black] (6,0) circle (2pt);
	\draw[very thick,red] (6,-1) -- (4,-1);
	\filldraw[red] (6,-1) circle (4pt);
	\filldraw[red] (4,-1) circle (4pt);
	\draw[very thick,blue] (0,-2) -- (2,-2);
	\filldraw[blue] (0,-2) circle (2pt);
	\filldraw[blue] (2,-2) circle (2pt);
\node[right,draw=black,rounded corners=3pt] at (-3.5,-4) {$S_{I_3}(A)=\ZZ_{u_2}+\xi(\ZZ_0-\ZZ_{u_1})$};
	\node[right] at (-3,0) {$A$};
	\node[right] at (-3,-1) {$\tilde I_3$};
	\node[right] at (-3,-2) {$A\ominus\tilde I_3$};
	\end{scope}
	\end{tikzpicture}
\end{center}
It is clear that $S(A)=S(R)$ as required by unitarity \cite{Hollowood:2021nlo}. 

Turning to the issue of decoupling, notice that the contributions from the island saddles $I_2$ and $I_3$ the entropy on $A$ factorizes, $S(A) = S(\overline R)+S(B_2)$, which automatically implies $I(\overline R,B_2)=0$.
But these saddles correspond to the condition that  $R$ is in one of its island saddles $I$ or $I'$. This motivates us to state the following island-in-the-stream decoupling rule: if two intervals are separated by a region large enough to be dominated by an island saddle, then they are uncorrelated.\footnote{This is related to the decoupling decoupling theorem IV.2 of \cite{ADHW}.}

It is easy to extend this result to prove that, on the other hand, two intervals are correlated, $I(A_1,A_2) >0$, if the entropy of $A_1 \cup A_2$ is dominated by an island saddle which image contains both $A_{1}$ and $A_2$, $A_1 \cup A_2 \subseteq \tilde{I}$.

\section{Entanglement monogamy and $A=R_B$}\label{s6}

In order for the Page curve \eqref{eq:Page_curve} to hold, there must be significant correlations between the early and late radiation. Since the Hawking radiation is also maximally entangled with its partners behind the horizon, including the reflected mode, we would have a violation of the entanglement monogamy if there is not some other mechanism at work (see the reviews \cite{Mathur:2009hf,Harlow:2014yka}). Thanks to entanglement-wedge reconstruction \cite{Jafferis:2015del,Dong:2016eik,Cotler:2017erl}, and its generalization to the island setting \cite{Almheiri:2020cfm}, it is now understood that the interior modes contained in an island are actually reconstructed in the radiation, realizing the so-called $A=R_B$ scenario, where $A$ are the interior partners (including modes reflected back into the black hole) of a subset of Hawking modes $B$ and $R_B$ is the purifier of $B$ in the Hawking radiation.\footnote{The notion of a purifier here is defined by the fact that the correlation measured by the mutual information is maximal between $B$ and $R_B$, $I(B,R_B)=2S(B)$ and vanishes with the remainder of the radiation $C$, $I(B,C)=0$. It is important to note that this notion of a purifier is not unique. This is indicative of the quantum error-correcting code interpretation of the correlations and the entanglement wedge reconstruction: the `information' is redundantly encoded in the radiation and is recoverable from different subsets thereof.}

\subsection{Correlations with the early and late radiation}

Take a small subset of the Hawking radiation $B=[u_1,u_2]$. Small in this context means that the entropy of $B$ is dominated by its Hawking saddle $S(B)=S_\text{rad}^{\mathbb T}(u_1,u_2)$. This means that $B$ cannot be too late, otherwise an island saddle dominates however small $B$ is. The early radiation is then $R_1=[0,u_1]$. The state of the black hole at time $u_2$ will then emerge in the late radiation $R_2=[u_2,u_\text{evap}]$.
It follows that we have a tripartite system in some overall pure state, where the entropy of each subsystem evaluated using \eqref{ger2} gives:
\begin{equation}
S(R_1)=\min\big(\xi(\ZZ_0-\ZZ_1),\ZZ_1\big) ,\quad S(B)=\xi(\ZZ_1-\ZZ_2) ,\quad S(R_2)=\min\big(\ZZ_2,\xi(\ZZ_0-\ZZ_2)\big) ,
\end{equation}
where $S(R_1)$ was computed in \eqref{eq:Page_curve} while $S(R_2)$ is given by a competition between saddles with islands-in-the-stream $\tilde I_1=R_2$ and $\tilde I_2=[0^-,u_\text{evap}]$. Since $B$ is a small interval, there are essentially two possibilities depending on whether $B$ is emitted before or after the Page time \eqref{nur}. Firstly, before the Page time $\ZZ_1>\ZZ_2>\ZZ_\text{Page}$, 
\EQ{
	S(R_1)=\xi(\ZZ_0-\ZZ_1)\ ,\quad S(R_2)=\xi(\ZZ_0-\ZZ_2)\ .
}
Since $B$ is a small subset of radiation of a young black hole, we have $S(R_1)+S(B)=S(R_2)=S(R_1\cup B)$. This implies that the mutual information $I(B,R_1)=0$ and so $B$ is not correlated with the early radiation $R_1$. On the other hand, it is maximally correlated with the late radiation $I(B,R_2)=2S(B)$. This is what we would expect from the interpretation of the Hawking radiation in terms of the Unruh effect, where outgoing Hawking modes, like the set $B$, are entangled with modes behind the horizon, modes that end up as the late radiation $R_2$ as the black hole evaporates. 

Then after the Page time we have $\ZZ_\text{Page}>\ZZ_1>\ZZ_2$ and therefore 
\EQ{
	S(R_1)=\ZZ_1\ ,\quad S(R_2)=\ZZ_2\ ,
}
and  we have the correlations 
\EQ{
	I(B,R_1)&=S(R_1)+S(B)-S(R_2)=\frac{\xi+1}\xi S(B)\ ,\\
	I(B,R_2)&=S(R_2)+S(B)-S(R_1)=\frac{\xi-1}\xi S(B)\ .
	\label{rix}
}
This adds nuance to the entanglement monogamy puzzle. For an old black hole, $B$ is not wholly entangled with either the early radiation $R_1$ or the remaining black hole, i.e.~what becomes the late radiation $R_2$. So a purifier of $B$ has to be partly in the early radiation and partly in the remaining black hole. The reversible case $\xi=1$ is special in that $I(B,R_1)=2S(B)$ and $I(B,R_2)=0$ and so a purifier of $B$ can lie wholly within the early radiation. 

An interesting issue is what happens when $B$ is chosen at a very late time so that $R_2$ shrinks to nothing. This seems not to be consistent with \eqref{rix}. The resolution is that \eqref{rix} assumes that $B$ is in its no-island saddle because it is a small interval. However, this cannot be maintained at very late time. Specifically, when $\xi(\ZZ_1-\ZZ_2)>\ZZ_1$, i.e.~$(\xi-1)\ZZ_1>\ZZ_2$, which is satisfied at late time as $R_2\to0$ ($\ZZ_2\to0$), $B$ switches to its island saddle. In that case, $I(B,R_1)\to2S(B)$.

\subsection{The purifier}

We have shown that if $B$ is emitted after the Page time, its purifier must be both in the early and late radiation. To explore this further, let us re-define $R_1=[u_1,u_2]$ and $R_2=[u_5,u_6]$ as smaller subsets of the early and late radiation with respect to $B=[u_3,u_4]$ and search for when $R_B=R_1\cup R_2$. In order to identify a purifier we have to require that $R_B$ and $B$ are maximally correlated $I(R_B,B)=2S(B)$. If the following island saddles dominate\footnote{If $u_1$ is early, i.e.~$\ZZ_1>\xi/(\xi+1)\ZZ_0$, then the island for $R_1$ will extend to $u=0^-$ but this does not affect the following arguments.}
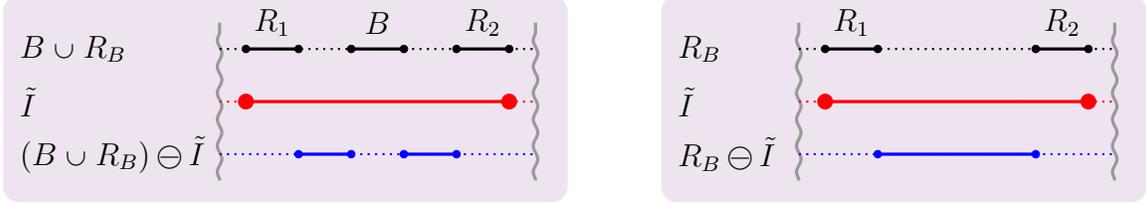
\begin{center}
	\begin{tikzpicture} [scale=0.7]
	\filldraw[fill = Plum!10!white, draw = Plum!10!white, rounded corners = 0.2cm] (-4.1,1) rectangle (6.6,-2.9);
	\draw[decorate,very thick,black!40,decoration={snake,amplitude=0.03cm}] (6,-2.5) -- (6,0.5);
	\draw[decorate,very thick,black!40,decoration={snake,amplitude=0.03cm}] (0,-2.5) -- (0,0.5);
	\draw[dotted,thick] (0,0) -- (6,0);
	\draw[dotted,thick,red] (0,-1) -- (6,-1);
	\draw[dotted,thick,blue] (0,-2) -- (6,-2);
	\draw[very thick] (2.5,0) -- (3.5,0);
	\draw[very thick] (0.5,0) -- (1.5,0);
	\draw[very thick] (4.5,0) -- (5.5,0);
	\filldraw[black] (0.5,0) circle (2pt);
	\filldraw[black] (1.5,0) circle (2pt);
	\filldraw[black] (2.5,0) circle (2pt);
	\filldraw[black] (3.5,0) circle (2pt);
	\filldraw[black] (4.5,0) circle (2pt);
	\filldraw[black] (5.5,0) circle (2pt);
	\draw[very thick,red] (0.5,-1) -- (5.5,-1);
	\filldraw[red] (0.5,-1) circle (4pt);
	\filldraw[red] (5.5,-1) circle (4pt);
	\draw[very thick,blue] (1.5,-2) -- (2.5,-2);
	\filldraw[blue] (1.5,-2) circle (2pt);
	\filldraw[blue] (2.5,-2) circle (2pt);
	\draw[very thick,blue] (3.5,-2) -- (4.5,-2);
	\filldraw[blue] (3.5,-2) circle (2pt);
	\filldraw[blue] (4.5,-2) circle (2pt);
	%\draw[very thick,blue] (2.5,-2) -- (3.5,-2);
	%\filldraw[blue] (2.5,-2) circle (2pt);
	%\filldraw[blue] (3.5,-2) circle (2pt);
	%
	\node at (1,0.5) {$R_1$};
	\node at (3,0.5) {$B$};
	\node at (5,0.5) {$R_2$};
	\node[right] at (-4,0) {$B\cup R_B$};
	\node[right] at (-4,-1) {$\tilde I$};
	\node[right] at (-4,-2) {$(B\cup R_B)\ominus\tilde I$};
	\begin{scope}[xshift=11cm]
	\filldraw[fill = Plum!10!white, draw = Plum!10!white, rounded corners = 0.2cm] (-2.6,1) rectangle (6.6,-2.9);
	\draw[decorate,very thick,black!40,decoration={snake,amplitude=0.03cm}] (6,-2.5) -- (6,0.5);
	\draw[decorate,very thick,black!40,decoration={snake,amplitude=0.03cm}] (0,-2.5) -- (0,0.5);
	\draw[dotted,thick] (0,0) -- (6,0);
	\draw[dotted,thick,red] (0,-1) -- (6,-1);
	\draw[dotted,thick,blue] (0,-2) -- (6,-2);
	\draw[very thick] (0.5,0) -- (1.5,0);
	\draw[very thick] (4.5,0) -- (5.5,0);
	\filldraw[black] (0.5,0) circle (2pt);
	\filldraw[black] (1.5,0) circle (2pt);
	\filldraw[black] (4.5,0) circle (2pt);
	\filldraw[black] (5.5,0) circle (2pt);
	\draw[very thick,red] (0.5,-1) -- (5.5,-1);
	\filldraw[red] (0.5,-1) circle (4pt);
	\filldraw[red] (5.5,-1) circle (4pt);
	\draw[very thick,blue] (1.5,-2) -- (4.5,-2);
	\filldraw[blue] (1.5,-2) circle (2pt);
	\filldraw[blue] (4.5,-2) circle (2pt);
	\node at (1,0.5) {$R_1$};
	\node at (5,0.5) {$R_2$};
	\node[right] at (-2.5,0) {$R_B$};
	\node[right] at (-2.5,-1) {$\tilde I$};
	\node[right] at (-2.5,-2) {$R_B\ominus\tilde I$};
	\end{scope}
	\end{tikzpicture}
\end{center}
\noindent it immediately follows that 
\EQ{
	S(R_B)-S(B\cup R_B)=S(B)\ ,
}
which implies maximal correlation:
\EQ{
	I(R_B,B)=2S(B)\ .
	\label{erg}
}

The constraints on $R_B=R_1\cup R_2$ that are needed for the islands above to dominate require that the intervals are not to small,
\EQ{
	S_\emptyset(R_1) \ge \frac{\xi + 1}{\xi-1} S_\text{rad}^{\mathbb T}(u_2,u_5) \ ,\qquad S_\emptyset (R_2) \ge \frac{\xi-1}{\xi+1} S_\text{rad}^{\mathbb T}(u_2,u_5)\ ,
	\label{vat}
}
and, as discussed in section \ref{s5.4}, we also need that the gap between $R_1$ and $R_2$ is small enough to be in its no-island saddle:
\EQ{
	S_\text{rad}^{\mathbb T}(u_2,u_5)<S_\text{BH}(u_2)+S_\text{BH}(u_5)\ .
	\label{rll}
}
These constraints do not uniquely define $R_B$ and so, as we alluded to earlier, there is freedom in defining a purifier.

Even if $R_B$ is maximally correlated with $B$, it is possible that for the individual intervals to have no  correlation with $B$, $I(B, R_1)=I(B,R_2)=0$. The seems surprising but is actually a typical feature of entanglement-wedge reconstruction and error-correcting codes \cite{Harlow:2016vwg,Almheiri:2014lwa}. We will show that it is only possible to achieve this situation in the irreversible case $\xi>1$.

The condition that $I(B,R_i)=0$ is equivalent to the requirement that the island-in-the stream of $R_i$ are not entropically favoured to extend to cover $B$ when we consider $R_i\cup B$. This requires
\EQ{
	R_1:\qquad&S_\text{rad}^{\mathbb T}(u_2,u_3)+S_\text{BH}(u_4)-S_\text{BH}(u_2)>S(B)\ ,\\[5pt]
	R_2:\qquad&S_\text{rad}^{\mathbb T}(u_4,u_5)+S_\text{BH}(u_3)-S_\text{BH}(u_5)>S(B)\ .
	\label{hec}
}
These conditions can be written as
\EQ{
	(\xi-1)\ZZ_2+(\xi+1)\ZZ_4>2\xi\ZZ_3\ ,\qquad (\xi-1)\ZZ_3+(\xi+1)\ZZ_5<2\xi\ZZ_4\ .
}
So if we write $\ZZ_4=\ZZ_3-\delta\ZZ$, for small $\delta\ZZ$, so $B=[\ZZ_3,\ZZ_3-\delta\ZZ]$, then we have
\EQ{
	\ZZ_2>\ZZ_3+\frac{\xi+1}{\xi-1}\delta\ZZ\ ,\qquad \ZZ_5<\ZZ_3-\frac{2\xi}{\xi+1}\delta\ZZ
}
and it is straightforward to see that \eqref{vat},\eqref{rll} are satisfied if we consider $R_{1,2}$ large enough, so $I(R_B,B)=2S(B)$. Note that this is only possible in the irreversible case because as $\xi\to1$, the first inequality cannot be satisfied whilst keeping $\ZZ_2<\ZZ_0$.

\section{Discussion}

In this paper we have derived a version of the generalized entropy for an arbitrary number of radiation intervals in the presence of a grey-body factor and in the adiabatic limit with intervals much larger than the scrambling time. The formula \eqref{ger2}, or \eqref{eq:gen3}, is a simple generalization of the `island-in-the-stream' formula first presented in \cite{Hollowood:2021nlo}. The fundamental parameter is $\xi$, which is the ratio of the transmitted entropy flux and the rate of decrease of the Bekenstein-Hawking entropy \eqref{rry}; the grey-body factor affects the formula by simply rescaling the radiation entropy by $\xi$. In JT gravity the grey-body factor is introduced by hand imposing partially transparent boundary conditions on the interface between the gravity and bath regions, which means that we can use it to control the evaporation rate. 

The generalized entropy has a drastically different behaviour in case the evaporation is reversible, $\xi = 1$, with respect to the irreversible one $\xi > 1$. Indeed, as discussed in section \eqref{s5.2}, multiple island saddles are not allowed in the reversible case, and the generalized entropy reduces to a competition of two saddles, as in Page's theorem.
The difference between reversible and irreversible evaporation appears also when studying the entanglement-monogamy problem, see section \ref{s6}. In both cases we manage to precisely identify a class of purifiers $R_B$ of a small radiation interval $B$ emitted after the Page time. However in the reversible case it turns out that $B$ can be entangled just with the early radiation, while in the irreversible case the purifier must be both in the past and the future of $B$.

The simple form of \eqref{eq:gen3} allows one deduce a lot of information-theoretical properties of the Hawking radiation. For example, it was proved in \cite{Hollowood:2021nlo} that all the entropy inequalities are automatically satisfied. In this spirit, in section \ref{s5.4} we proved a decoupling theorem, which can be summarized in the statement that two radiation intervals $R_{1,2}$ are not correlated, $I(R_1,R_2)=0$, when the interval in between is dominated by an island saddle.

One important point shown in section \ref{s5} is that also the irreversible islands-in-the-stream formalism can be captured by a remarkably simple quantum statistical mechanics model involving a temporal sequence of unitary averages. One can view this as a nested generalization of Page's theorem. The fact that ensemble averages are involved is not a surprise because there are strong arguments that the semi-classical gravitational path integral is actually computing an average over some more microscopic description \cite{Pollack:2020gfa,Cotler:2020ugk,Bousso:2020kmy}. This is immediately clear in JT gravity \cite{Saad:2019lba,Stanford:2019vob} but has been argued generally in \cite{Marolf:2020rpm}. In this latter work the ensemble average arises from the fact that when computing a R\'enyi entropy using the replica method, there is something special about a gravitational theory because the way that replicas are joined along a future Cauchy surface can itself fluctuate. The upshot is that the Lorentzian replica wormhole and associated island correspond to a region on the Cauchy slice where the replicas are joined in a non-trivial way. This kind of occurrence is familiar from the theory of baby universes and an ensemble interpretation arises in the same way. It would be interesting to relate this baby universe ensemble to the sequence of unitary averages that we have found describing the correlations in the Hawking radiation.

One way to interpret the unitary averages in our model is as in statistical mechanics, so the ensemble captures the time-averaged behaviour of a single system. The idea is that over a time scale that is much greater than  the scrambling time the black hole equilibrates \cite{Sasieta:2021pzj,Krishnan:2021faa}. This means that various coarse-grained observables like entropies are approximately equal to the same observables in some appropriate equilibrium ensemble. So effectively we can replace the time-evolution of some particular state $\rho$ by the appropriate equilibrium state $\rho_\text{eq}$. When computing R\'enyi entropies the replacement of the replica state takes the form of the equilibration ansatz of \cite{Liu:2020jsv}:\footnote{More precisely the simplified form which will be valid here since $N$ is large and the equilibrium state is maximally mixed.} 
\EQ{
U_{\delta t}^{\otimes n}\rho^{\otimes n}U_{\delta t}^{\dagger\otimes n}\longrightarrow \frac1{Z_2^n}\sum_{\tau\in S_n}\tr^{(n)}(\tau\rho^{\otimes n}\rho_\text{eq}^{\otimes n})\tau\rho_\text{eq}^{\otimes n}\ ,
\label{evf}
}
where $\delta t\gg\Delta t_s$ and the normalization factor $Z_2=\tr(\rho_\text{eq}^2)$. In the case that the relevant equilibrium state is the maximally mixed state $\rho_\text{eq}={\bf 1}/N$, the right-hand side of \eqref{evf} is precisely the same as the unitary average \eqref{vuu}. It follows that the equilibration ansatz of \cite{Liu:2020jsv} leads to the same expressions for the entropy as we have found.

\vspace{0.5cm}
\begin{center}{\it Acknowledgments}\end{center}
\vspace{0.2cm}
TJH, AL and SPK acknowledge support from STFC grant ST/T000813/1. NT acknowledges the support of an STFC Studentship

\appendix
\appendixpage

\section{JT black hole and back reaction}\label{a3}

For the JT gravity model coupled to massless scalars, the back reaction problem can be solved exactly in the semi-classical limit. The simplicity of the model lies in the fact that the metric is fixed to be AdS$_2$ which is glued onto a half Minkowski space near spatial infinity. The AdS$_2$ part has a metric in KS coordinates
\EQ{
ds^2=-\frac{dU\,dV}{(1+UV)^2}\ .
}
The non-trivial aspect of the model is how this is glued onto the half-Minkowksi space $ds^2=-du\,dv$, $v-u>0$, along $u=v$. This is described by a function $\sigma(t)$:
\EQ{
U=-e^{-\sigma(u)}\ ,\qquad V=e^{\sigma(v)}\ .
\label{rcc}
}
So in a way that will emerge, JT gravity boils down to a theory of this one function $\sigma(t)$.

When JT gravity is coupled to matter fields, the dilaton becomes dynamical and satisfies the following equations where the matter stress tensor acts as a source:
\EQ{
-\frac1{(1+UV)^2}\partial_U\big((1+UV)^2\partial_U\phi\big)&=8\pi G_NT_{UU}\ ,\\[5pt]
\partial_U\partial_V\phi+\frac2{(1+UV)^2}(\phi-\phi_*)&=8\pi G_NT_{UV}\ ,\\[5pt]
-\frac1{(1+UV)^2}\partial_V\big((1+UV)^2\partial_V\phi\big)&=8\pi G_NT_{VV}\ .
}
In the case at hand, the matter consists of $\mathcal{N}$ massless scalar fields. The outgoing modes are assumed to be in the $U$ vacuum, so that $T_{UU}=0$ and also $T_{UV}=0$. However, when there are partially reflecting boundary conditions there will be an infalling component $T_{VV}$. To this end, we solve the above with $T_{VV}\neq0$. The first two equations determine
\EQ{
\phi=\phi_*+\frac{\partial_VF(V)}2-\frac{UF(V)}{1+UV}\ .
}
The third equation determines the function $F$:
\EQ{
-\frac12\partial_V^3F=8\pi G_NT_{VV}\ .
\label{hue}
}

The other input is the energy conservation at the interface between the AdS$_2$ and Minkowski regions. We identify the mass of the black hole with the ADM energy plus the extremal mass:
\EQ{
M=M_* -\frac{\phi_r}{8\pi G_N}\{e^\sigma,t\}\ .
}
Here, $t=u=v$ is the time at the boundary and 
\EQ{
\{f,t\}=\frac{\dddot f }{\dot f}-\frac32\Big(\frac{\ddot f}{\dot f}\Big)^2\ ,
} 
is the Schwarzian. The energy conservation condition is then
\EQ{
\frac{dM}{dt}=\big(T_{vv}-T_{uu}\big)\Big|_{u=v=t}\ .
\label{nut}
}
Here, $T_{vv}$ is the stress tensor of the modes reflected off the boundary. 

The outgoing modes are in the $U$ vacuum, i.e.~$T_{UU}=0$ and therefore we can make a conformal transformation $U\to u$ to get
\EQ{
T_{uu}=-\frac{\cal N}{24\pi}\{e^\sigma,u\}\ .
\label{nat}
}
On the other hand, for the infalling sector, we have
\EQ{
T_{VV}=(\dot\sigma e^\sigma)^{-2}\Big(T_{vv}+\frac{\cal N}{24\pi}\{e^\sigma,v\}\Big)\ .
}
Given \eqref{nat}, and replacing $u\to t$, and substituting into \eqref{nut} and the replacing $t\to v$ gives
\EQ{
T_{vv}+\frac{\cal N}{24\pi}\{e^\sigma,v\}=\frac{dM(v)}{dv}\ .
}
Hence, 
\EQ{
T_{VV}=(\dot\sigma e^\sigma)^{-2}\dot M=-\frac{\phi_r}{8\pi G_N}(\dot\sigma e^\sigma)^{-2}\partial_v\{e^\sigma,v\}\ .
}
Since $\dot M<0$, $T_{VV}<0$ and therefore there is a flux of negative energy across the horizon. Of course this is needed to balance the net flux of Hawking radiation away from the black hole. Now we can change the $v$ derivatives into $V$ derivatives to get
\EQ{
T_{VV}=-\frac{\phi_r}{8\pi G_N}\partial_V^3(\dot\sigma e^{\sigma})\ .
}
Comparing this to \eqref{hue}, determines $F$ in terms of $\sigma$, up to some integration constants that can be absorbed into the definition of $\sigma$:
\EQ{
F(e^\sigma)=\frac{\phi_r\dot\sigma e^{\sigma}}{4\pi G_N}\ . 
}
So the dilaton is also determined by the function $\sigma$.

So the remaining problem is to solve for $\sigma$ from the energy conservation equation
\EQ{
\frac{dM}{dt}=-\frac{\phi_r}{8\pi G_N}\partial_t\{e^\sigma,t\}=T_{vv}(t)+\frac{\cal N}{24\pi}\{e^\sigma,t\}\ ,
\label{ysi}
}
given the particular boundary conditions imposed. 

Our boundary conditions are given at the level of the modes by \eqref{het2} with the transmission and reflection coefficients $\mathbb R(\omega)$ and $\mathbb T(\omega)$ as input. Rather than solve for $T_{vv}$ for the reflected modes in general, we are primarily interested in the adiabatic limit which we will use to simplify the problem. We have shown that the outgoing stress tensor component $T_{uu}$ in \eqref{nat} involves the Schwarzian of $e^\sigma$:
\EQ{
\{e^\sigma,t\}=-\frac12\dot\sigma^2-\frac32\Big(\frac{\ddot\sigma}{\dot\sigma}\Big)^2+\frac{\dddot\sigma}{\dot\sigma}\ .
}
In the adiabatic approximation, we keep only the first term here, and so
\EQ{
T_{uu}\approx\frac{\cal N}{48\pi}\dot\sigma^2\ .
\label{xer}
}

\subsection{Identifying the temperature}

There are two ways to identify the slowly varying temperature of the evaporating hole. Firstly, we can use the thermodynamic relation \eqref{bur}. In the adiabatic limit,
\EQ{
\dot M\approx\frac{\phi_r}{8\pi G_N}\dot\sigma\ddot\sigma
}
and the Bekenstein-Hawking entropy is
\EQ{
\SBH=S_*+\frac{\phi(U=0)}{4G_N}=S_*+\frac{\partial_VF}{8G_N}\ .
}
Hence,
\EQ{
\dot S_\text{BH}=\frac1{8G_N}\partial_v\partial_V(2\phi_r\dot\sigma e^\sigma)\approx \frac{\phi_r}{4G_N}\ddot\sigma
}
and so 
\EQ{
T=\frac{\dot\sigma}{2\pi}\ .
\label{muy}
}
which gives rise the expressions \eqref{ruz}.

The temperature also follows from the usual Bogoliubov transformation approach of Hawking's original calculation carefully generalized to account for the time-dependent temperature \cite{Barcelo:2010xk}. 
The relevant quantity is the Bogoliubov coefficient for the transformation between a positive frequency mode $e^{-i\omega' U}$ and a negative frequency mode $e^{i\omega u}$, $\omega,\omega'>0$. In particular, using conventional notation this coefficient is
\EQ{
\beta=\frac1{2\pi}\sqrt{\frac{\omega}{\omega'}}\int_{-\infty}^\infty du\,e^{-i\omega u-i\omega' U}=\frac1{2\pi}\sqrt{\frac{\omega}{\omega'}}\int_{-\infty}^\infty du\,\exp\big[-i\omega u+i\omega' e^{-\sigma(u)}\big]\ .
\label{pep}
}
In the original calculation, the temperature is taken to be constant and $\sigma(u)=2\pi Tu$, but when it varies the situation is more subtle. The rigorous way to proceed is to work with wave packets that are localized in some window $u\in[u_*-\Delta u,u_*+\Delta u]$ effectively restricting the integral to the window \cite{Barcelo:2010xk}. In the adiabatic limit, where the variation of the temperature occurs on much larger scales than $\Delta u$, we can approximate $\sigma(u)$ inside the integral by its expansion to first order around $u_*$,
\EQ{
\sigma(u)=\sigma_*+(u-u_*)\,\dot\sigma_*\ ,
}
where $\sigma_*=\sigma(u_*)$, in the integral in \eqref{pep}. Then the limits of the integral can be extended to $\pm\infty$ and be evaluated by the saddle point method (or exactly) leading to an equation
\EQ{
\omega=-\omega'\dot\sigma_*e^{-\sigma_*-(u-u_*)\,\dot\sigma_*}\ .
}
The solution has an imaginary part 
\EQ{
u=u_*-\frac{i\pi}{\dot\sigma_*}-\frac1{\dot\sigma_*}\Big(\sigma_*-\log\frac{\omega'\dot\sigma_*}\omega\Big)\ .
}
The imaginary part means that $|\beta|^2$ has an exponential factor
\EQ{
|\beta|^2=\frac1{2\pi\sigma_*\omega'}e^{-2\pi\omega/\dot\sigma_*}\ ,
}
which becomes the Boltzmann factor $e^{-\omega/T}$ and so identifies the temperature at $u=u_*$ as \eqref{muy} above. 

\subsection{Solving for $T(t)$}

Our focus is now on solving \eqref{ysi} in the adiabatic limit. Given \eqref{xer} we have
\EQ{
T_{uu}=\frac{\pi cT^2}6={\cal N}\int_{-\infty}^\infty\frac{d\omega}{2\pi}\cdot\frac{\omega}{e^{\omega/T}-1}\ .
}
which is precisely the expression expected for a relativistic bosonic gas. We can put in the stress tensor of the reflected modes $T_{vv}$. This will be the energy flux of the bosonic gas with the reflection factor:
\EQ{
T_{vv}={\cal N}\int_{-\infty}^\infty\frac{d\omega}{2\pi}\cdot\frac{\omega(1-\Gamma(\omega))}{e^{\omega/T}-1}\ .
}
Hence, \eqref{ysi} becomes the evaporation equation \eqref{her}. For the near-extremal black hole, the mass is
\EQ{
M=M_*-\frac{\phi_r}{8\pi G_N}\{e^\sigma,t\}\approx M_*+\frac{\phi_r}{16\pi G_N}\dot\sigma^2=M_*+\frac{\pi\phi_rT^2}{4G_N}\ .
}

\section{Schwarzschild black hole and back-reaction}\label{a4}

The back-reaction of Hawking radiation on a Schwarzschild black hole and its evaporation is a problem that has not be solved exactly. However, there is an approximate understanding valid in the adiabatic limit when the evaporation is sufficiently slow \cite{Bardeen:1981zz,Parentani:1994ij,Massar:1994iy,Abdolrahimi:2016emo}. What these references establish is that there is a consistent way to correct the Schwarzschild metric to take into the back-reaction of Hawking radiation. It is not the purpose of this appendix to review the problem in complete detail but we will review sufficient details to establish the results that we need in the text.

As the black hole evaporates, the black hole can be assigned a time-dependent mass $M(t)$ which satisfies\footnote{In this appendix we choose units for which $G_N=1$ (as well as $c=\hbar=1$).}
\EQ{
\frac{dM}{dt}=-L_H=-\frac C{M^2}\ ,\qquad C=\frac{{\cal N}\eta}{3\cdot2^8\pi}\ ,
\label{rer}
}
where $L_H$ is the flux of Hawking radiation. Clearly in the adiabatic limit \eqref{hrr}, that is $M^2\gg{\cal N}$, the flux is small $L_H\ll1$. There are essentially two different definitions of the mass for the evaporating black hole: firstly in terms of the apparent horizon and secondly in terms of the fall off at large $r$. In general these definitions will differ, but in the adiabatic limit $L_H\ll1$, this difference will be very small.

Specifically we will establish the following, to leading order in the adiabatic approximation:
\begin{enumerate}
\item In the neighbourhood of a sufficiently small region of the horizon, there exist Kuskal-Szekeres type coordinates $(U,V)$ for which the radial coordinate has the expansion
\EQ{
r=r_h(v)(1-UV+\cdots)\ ,
\label{ozn}
}
where $V=V(v)$. In the above, $r_h(v)\approx 2M(v)$ is the position of the horizon, i.e.~$U=0$, where $v$ is an in-going Eddington-Finkelstein type coordinate. This implies that the area of a 2-sphere in the neighbourhood of the horizon is
\EQ{
\text{Area}(S^2)=4S_\text{BH}(v)\big(1-2UV+\cdots\big)\ .
}
to linear order in $U$, where $S_\text{BH}(v)=\pi r_h(v)^2$.
\item In terms of the KS-type coordinates, the near-horizon metric is
\EQ{
ds^2\Big|_\text{near hor.}\approx-16M(v)^2\,dV\,dU+4M(v)^2(1-2UV)d\Omega^2
\label{gff}
}
and so $U$ is an inertial coordinate. 
\item To leading order, in the near horizon zone
\EQ{
\frac{dV}{dv}\approx \frac V{4M(v)}\ ,\qquad
\frac{dU}{du}\approx-\frac U{4M(u)}\ ,
\label{faf}
}
where $u$ and $v$, the inertial coordinates far from the black hole, are shifted so that the endpoint of the evaporation occurs at the same time $u_\text{evap}$. The Jacobian between $U$ and $u$ ensures that the black hole radiates quasi-thermally with a slowly varying temperature $T(u)\approx (8\pi M(u))^{-1}$. Note that $u=v$, i.e.~$r_*=0$, occurs near the edge of the zone.
\end{enumerate}

The analyses of \cite{Massar:1994iy,Abdolrahimi:2016emo} use a parameterization of a general spherically symmetric metric in terms of in-going Eddington-Finkelstein type coordinates $(v,r)$ of the form:
\EQ{
ds^2=-e^{2\psi(v,r)}\Big(1-\frac{2m(v,r)}r\Big)dv^2+2e^{\psi(v,r)}dv\,dr+r^2d\Omega^2\ ,
\label{tiu}
}
with $\psi \to 0$ as we approach the near horizon region \cite{Massar:1994iy}.
The quantity $m(v,r)$ determines the apparent horizon via the solution of $r_a(v)=2m(v,r_a(v))$ which in turn providing a definition of the mass of the black hole $M(v)=r_a(v)/2$ \cite{Bardeen:1981zz}. It is shown in \cite{Massar:1994iy} that $M(v)$ approximately satisfies \eqref{rer} with the identification $v=t$ (with an appropriate choice of origin for $v$). The apparent horizon is not the same as the event horizon which is defined as the last outgoing null ray that reaches $\mathscr I^+$. It is determined by the solution to the equation for an outgoing null geodesic
\EQ{
\frac{dr_h}{dv}=\frac{r_h-2m(v,r_h)}{2r_h}\ .
\label{lik1}
}
This determines $r_h=2M-8C/M+\cdots$ and so is slightly smaller than the apparent horizon, however, to  leading order $r_h(v)\approx 2M(v)$. 

Further out at the edge of the `zone' for Hawking modes, say around $r$ $\sim 6M$, just beyond the potential barrier, it is consistent to write the metric as the Vaidya metric 
\EQ{
ds^2=-\Big(1-\frac{2m(u)}r\Big)du^2-2du\,dr+r^2d\Omega^2\ .
\label{vad}
}
Outside the zone, there is just an outgoing flux of Hawking radiation and a corresponding stress tensor $T_{uu}=-L_H(u)/4\pi r^2$ and $dm(u)/du=-L_H(u)$. Hence, the new mass function $m(u)$ also satisfies \eqref{rer}, so to leading order in the adiabatic approximation, we can identify $m(u)\approx M(u)$ with a suitable choice of origin for $u$. 

The change of coordinates $(v,r)$ to $(u,r)$ involves
\EQ{
e^\psi dv=du+\frac{2\,dr}{1-2m(u)/r}
\label{lic}
}
and $m(v,r)=m(u)$. The mass functions $m(v,r)$ and $m(u)$ are slowly varying functions with gradients suppressed by the Hawking flux $L_H$ \cite{Massar:1994iy}. 

Notice that the mass function $M(t)$ has now appeared in two different ways: as function of $v$ as the position of the apparent horizon $2M(v)$ and as a function of $u$ via the fall off at large radius of the Vaidya metric $M(u)$. These two definitions are consistent in the region $u\sim v$ as we move from the metric inside the zone to the one outside (assuming both origins have been fixed so that the end of the evaporation occurs at $u=v$).

The metric in the neighbourhood of the horizon, outside and slightly inside the black hole takes the form \eqref{tiu}. In this region, we can ignore $\psi$ and make the coordinate transformation $r=r_h(v)+z\approx 2M(v)+z$, giving 
\EQ{
ds^2\approx-\frac{z}{2M(v)+z}\,dv^2+2dv\,dz+4M(v)^2\Big(1+\frac z{M(v)}\Big)d\Omega^2\ .
\label{tiu2}
}
In the neighbourhood of the horizon, i.e.~small $z$, let us change variable from $z$ to an inertial coordinate $ U=-e^{-\alpha(v)}z$ for some $\alpha(v)$ which is a solution of 
\EQ{
\frac{d\alpha(v)}{dv}=\frac{1}{4M(v)}\ .
\label{lik2}
}
This gives 
\EQ{
ds^2\Big|_\text{near hor.}\approx -2e^{\alpha(v)}dv\,dU+4M(v)^2\Big(1-U\frac{e^{\alpha(v)}}{M(v)}\Big)d\Omega^2\ ,
}
which shows that $U$ is, indeed, inertial.

The next step is to define $V(v)$ according to \eqref{ozn}, yielding 
\EQ{
V=\frac{e^{\alpha(v)}}{2M(v)}\ .
}
Using \eqref{lik1} and \eqref{lik2} and ignoring derivatives of $M(v)$, we have \eqref{faf}
\EQ{
\frac{dV}{dv}\approx\frac{e^{\alpha(v)}}{8M(v)^2}=\frac V{4M(v)}\ .
\label{gdd}
}
The near-horizon metric then takes the advertized form \eqref{gff}.

The final result we need is \eqref{faf}. From the change of variables between $(v,r)$ and $(u,r)$ in \eqref{lic}, we have
\EQ{
\frac{du}{dr}=-\frac{2r}{r-2m(u)}\ .
}
In the near-horizon zone $r=2M(u)+z$, with $M(u)\equiv m(u)$.
Therefore, near the horizon we have $du/dr\approx-4M(u)/z$; hence, at constant $v$, and using \eqref{faf},
\EQ{
\frac{dU}{du}=\frac{dU}{dr}\frac{dr}{du}=e^{-\alpha}\frac{z}{4M(u)}=-\frac{U}{4M(u)}\ ,
}
which is \eqref{faf}.

So to compare with the expressions in the JT gravity model, the near-horizon relations between the coordinates $(U,V)$ and $(u,v)$ are the same; compare \eqref{rcc} with \eqref{faf} above, given that $M(t)=(8\pi T(t))^{-1}$. The only difference is that the near-horizon metric \eqref{gff} has the conformal factor $16M(v)^2$. However, this does not affect any of our conclusions as we point out in the text. Such a factor would in any case be present for the Reissner-Nordstr\" om black hole in $3+1$ under the $s$-wave reduction to JT gravity.

\end{document}